%% file: main.tex
\documentclass[
aps,
twocolumn,
floatfix,
10pt,
superscriptaddress,
longbibliography
]{revtex4-2}
\input{PackagesAndSettings}

\begin{document}

\title{Quasicrystalline Analogue of the Haldane Model}

\author{B. D. K. Burgess}
\email{bb611@cam.ac.uk}
\affiliation{TCM Group, Cavendish Laboratory, University of Cambridge, J.J. Thomson Avenue, Cambridge, CB3 0US, United Kingdom\looseness=-1}

\author{N. R. Cooper}
\affiliation{TCM Group, Cavendish Laboratory, University of Cambridge, J.J. Thomson Avenue, Cambridge, CB3 0US, United Kingdom\looseness=-1}

\begin{abstract}
    We present a model for a topological quasicrystalline system which is suitable for realisation in cold atom experiments. We define the model in terms of complex momentum-space couplings which break time-reversal symmetry (TRS), and detail how it may be experimentally realised using two-photon Raman couplings. In the weak-potential limit, we study the model analytically by calculating the bandstructure over a `quasi-Brillouin zone' (QBZ). We find symmetry-protected Dirac cones, which are gapped by a TRS-breaking term, resulting in a Chern number $\mathcal{C}=1$. This provides a direct analogy to the Haldane model, but now in a quasicrystalline setting. We also infer the number of states below the topological gap from the QBZ area. We verify our analysis with numerical calculations of periodic approximants to our system, constructing a phase diagram in parameter space which shows a topological region extending beyond the weak-potential regime. We also find examples of narrow Chern bands with the potential for hosting strongly-correlated physics. Our work raises questions about the nature of localisation and strongly-correlated states in Chern bands in quasiperiodic systems.
\end{abstract}

\maketitle

\input{Introduction}
\input{Model}
\input{Experimental}
\input{PlaneWave}
\input{Approximant}
\input{Summary}

\begin{acknowledgments}
We thank U. Schneider and Y. Yang for helpful discussions. This work was supported by the EPSRC [Grant No. EP/V062654/1 and EP/Y01510X/1] and a Simons Investigator Award [Grant No. 511029].
\end{acknowledgments}

\section*{Data Availability}
The data and code supporting the findings of this article are openly available \cite{Data}.

\appendix
\input{Pentagonal}
\input{Coherent}
\input{Symmetries}
\input{Trends}

\bibliography{refs}

\end{document}

%% file: PackagesAndSettings.tex
\usepackage{silence}
\usepackage[utf8]{inputenc}
\usepackage[T1]{fontenc}
\usepackage{textcomp}
\usepackage{newunicodechar}
\newunicodechar{∘}{\textdegree}
\usepackage{lmodern}
\usepackage{amsmath,amssymb,amsthm}
\usepackage{physics}
\usepackage{graphicx}
\usepackage{hyperref}
\usepackage{xcolor}
\usepackage{braket}
\usepackage{bm}
\usepackage{bbm}
\usepackage{csquotes}
\usepackage{floatrow}
\usepackage[labelformat=simple]{subfig}
\usepackage{ragged2e}
\makeatletter
\long\def\@makecaption#1#2{%
  \vskip\abovecaptionskip
  \sbox\@tempboxa{\justifying\small #1: #2}%
  \ifdim \wd\@tempboxa >\hsize
    {\justifying\small #1: #2\par}
  \else
    \global \@minipagefalse
    \hb@xt@\hsize{\hfil\box\@tempboxa\hfil}%
  \fi
  \vskip\belowcaptionskip}
\makeatother

\hypersetup{
    colorlinks=true,
    linkcolor=blue,
    citecolor=blue,
    urlcolor=blue
}

\definecolor{revision}{RGB}{80, 160, 80}

%% file: Introduction.tex
\section{Introduction}\label{Sec:Intro}

Quasicrystals are spatially ordered but not periodic \cite{Shechtman_Metallic_1984, Lifshitz_Quasicrystals_2003}, making them a middle-ground between aperiodic and crystalline systems. In recent years, quasicrystals have been of particular interest in their unconventional localisation properties of quantum particles, which differ from those of randomly disordered systems. Although all single-particle states are Anderson-localised by random disorder in spatial dimension $d\!\leq\!2$ \cite{anderson_absence_1958, abrahams_scaling_1979, evers_anderson_2008}, many-body localisation (MBL) of interacting particles is theorised to be impossible in $d\!=\!2$ due to rare ergodic regions which trigger an `avalanche of thermalisation' \cite{de_roeck_stability_2017, luitz_how_2017, thiery_many-body_2018, potirniche_exploration_2019, morningstar_avalanches_2022, sels_bath-induced_2022, leonard_probing_2023}. Quasicrystals provide a form of `deterministic disorder' with no rare regions, and are therefore widely researched with the goal of realising MBL in $d\!=\!2$ \cite{szabo_mixed_2020, strkalj_coexistence_2022, Thomson_Unravelling_2024, bordia_probing_2017, Sbroscia_Observing_2020, hur_stability_2025}. Quasicrystals in $d\!=\!1$ also display interesting properties such as mixed spectra where extended and localised states coexist \cite{biddle_predicted_2010, biddle_localization_2011, ganeshan_nearest_2015, li_mobility_2020, an_interactions_2021, wang_observation_2022, devakul_anderson_2017, szabo_mixed_2020, Gottlob_Hubbard_2023}, or intermediate `critical' states \cite{Zhu_Localization_2024, duncan_critical_2024}.

When the energy states of a quasicrystal can also be characterised by some topological invariant, even richer effects can occur due to the interplay between the quasiperiodicity and the topology \cite{Fan_Topological_2021, Zilberberg_Topology_2021}. Recently, such topological quasicrystals have been widely investigated through the numerical analysis of real-space tight-binding models on quasiperiodic lattices such as the Penrose or Ammann-Beenker tilings. A wide range of phenomena have been reported, including unconventional `bulk-localised transport' states \cite{Duncan_Topological_2020, Johnstone_Bulk_2022, balling-anso_identification_2025}, higher-order topological insulators \cite{hua_higher-order_2020, Chen_HigherOrder_2020, Peng_Higherorder_2021, Peng_Structural_2024, Varjas_Topological_2019, Wang_Effective_2022, Yang_Higherorder_2024}, topological pumping and protected transport \cite{Bandres_Topological_2016, Gottlob_Quasiperiodicity_2025, He_Quasicrystalline_2019, Huang_Quantum_2018, Kraus_Topological_2012, Kraus_Topological_2012a, Kraus_FourDimensional_2013, peng_topological_2021, Petrides_Sixdimensional_2018, Tran_Topological_2015, Ghadimi_Confined_2023}, and topological superconductivity \cite{DeGottardi_Majorana_2013, Fulga_Aperiodic_2016, Ghadimi_Majorana_2017, Ghadimi_Topological_2021, Tezuka_Reentrant_2012, Cao_KohnLuttinger_2020, Manna_Noncrystalline_2024}. In contrast, experimental research into topological quasicrystals is limited to a small number of photonic experiments \cite{Kraus_Topological_2012, verbin_observation_2013, Verbin_Topological_2015,Che_Polarization_2021}, which are well-suited to implementing such tight-binding models.

Ultracold atoms in optical lattices provide an alternative and appealing experimental setting for investigating topological quasicrystals. These systems have emerged as a leading platform for quantum simulation due to their exceptional degree of tunability \cite{Bloch_Quantum_2012, Gross_Quantum_2017, Schafer_Tools_2020}. Previous cold atom experiments have studied quasicrystalline \cite{Roati_Anderson_2008, Schreiber_Observation_2015, Singh_Fibonacci_2015, Viebahn_MatterWave_2019, Sbroscia_Observing_2020, Guidoni_Quasiperiodic_1997} and topological \cite{Jaksch_Creation_2003, Aidelsburger_Realization_2013, Miyake_Realizing_2013, Jotzu_Experimental_2014, Cooper_Topological_2019} systems independently, raising the natural question of whether they can be employed to study both together. However, to the best of our knowledge, no proposals exist for realizing tight-binding models of the type studied in Refs. \cite{Duncan_Topological_2020, Johnstone_Bulk_2022, balling-anso_identification_2025, hua_higher-order_2020, Chen_HigherOrder_2020, Peng_Higherorder_2021, Peng_Structural_2024, Varjas_Topological_2019, Wang_Effective_2022, Bandres_Topological_2016, Gottlob_Quasiperiodicity_2025, He_Quasicrystalline_2019, Huang_Quantum_2018, Kraus_Topological_2012, Kraus_Topological_2012a, Kraus_FourDimensional_2013, peng_topological_2021, Petrides_Sixdimensional_2018, Tran_Topological_2015, DeGottardi_Majorana_2013, Fulga_Aperiodic_2016, Ghadimi_Majorana_2017, Ghadimi_Topological_2021, Tezuka_Reentrant_2012, Cao_KohnLuttinger_2020, Ghadimi_Confined_2023, Manna_Noncrystalline_2024, Yang_Higherorder_2024} with cold atoms:  engineering the required complex hopping parameters on a quasiperiodic optical lattice is an open problem.

In this work, we instead present a continuum model for a topological quasicrystal, which is formulated directly in reciprocal space and naturally suited to cold atom experiments. Specifically, we construct a quasiperiodic version of the `optical flux lattice' (OFL) scheme for generating topological bandstructures \cite{Dalibard_Colloquium_2011, Cooper_Optical_2011, Cooper_Optical_2011a, Cooper_Designing_2012, Cooper_Topological_2019}. We will show that the model possesses distinctive symmetry and topological properties, with close parallels to the Haldane model~\cite{haldane_model_1988} despite the absence of spatial periodicity. Although motivated by cold atoms, we note that our model may also be realisable in 2D layered materials, where quasicrystalline systems \cite{ahn_dirac_2018, yu_dodecagonal_2019, moon_quasicrystalline_2019, crosse_quasicrystalline_2021, layered-quasicrystal} and (periodic) OFL Hamiltonians \cite{Wu2019, Sommer_Ideal_2025} are known to emerge. However, we stress that the model stands independently of any specific implementation and is of theoretical interest in its own right, being distinct from the real-space tight-binding models that have dominated previous works. 

The remainder of the paper is structured as follows. We motivate our model Hamiltonian with reference to previous work on OFLs in Sec. \ref{Sec:Model}; we then detail exactly how it may be experimentally realised in Sec. \ref{Sec:Experimental}. Next, we analyse the model in Secs. \ref{Sec:PlaneWave}-\ref{Sec:Approximant}: we demonstrate that it is topologically non-trivial, construct the phase diagram of the system in parameter-space, and investigate the particle density required to fill all states below the topological energy gap. We summarise our work in Sec. \ref{Sec:Summary}.

%% file: Model.tex
\section{Model}\label{Sec:Model}

\subsection{Optical Flux Lattices}

We first briefly introduce OFLs in general terms, to explain how these give rise to topological bandstructures. For more detailed discussions and derivations of the results quoted here, we direct the reader to the reviews of Refs. \cite{Dalibard_Colloquium_2011, Cooper_Topological_2019}. We consider a particle with $N$ internal `spin' states moving in 2D, with Hamiltonian:
\begin{equation}\label{Eq:GeneralHamiltonian}
    H = \frac{\mathbf{p}^2}{2m}\otimes\mathbbm{1}_N + V(\mathbf{r})\, ,
\end{equation}
where $V(\mathbf{r})$ is an $N\times N$ matrix coupling different spin states. In the adiabatic limit of large potential energy, the particle follows the same local spinor eigenstate $\ket{n(\mathbf{r})}$ of $V(\mathbf{r})$. As it moves through space, it acquires a Berry phase due to the changing of $\ket{n(\mathbf{r})}$ with position. By projecting $H$ onto the spinor texture $\ket{n(\mathbf{r})}$, one finds an effective Hamiltonian which is that of a scalar particle moving in the background of a gauge potential \cite{Cooper_Optical_2011}:
\begin{equation}
    \mathbf{A}(\mathbf{r}) = \textrm{i}\braket{n(\mathbf{r})|\nabla_\mathbf{r}n(\mathbf{r})}
\end{equation}
with associated `effective magnetic field':
\begin{equation}
    B_{\text{eff}} (\mathbf{r}) = \nabla_\mathbf{r}\times \left[\textrm{i} \braket{n(\mathbf{r})|\nabla_\mathbf{r}n(\mathbf{r})}\right]\cdot\hat{\mathbf{z}}\,.
\end{equation}
(We set $\hbar=1$.) These are respectively the (real-space) Berry potential and Berry curvature of $\ket{n(\mathbf{r})}$. This emergent magnetic field explains why OFLs acquire non-trivial topological properties for a suitable form of $V(\mathbf{r})$. 

It is also instructive to consider a reciprocal-space description of OFLs \cite{Cooper_Designing_2012}. The potential energy term has Fourier components:
\begin{equation}
    V_{\mathbf{G},\,\sigma^\prime\sigma} = 
    \braket{\mathbf{q}+\mathbf{G}, \sigma^\prime|V(\mathbf{r})|\mathbf{q},\sigma}\,,
\end{equation}
where $\mathbf{q}$ is the particle momentum, and $\sigma$ indexes the spin states. Again working in the limit of zero kinetic energy, the Hamiltonian takes the form of a tight-binding model, where $V_{\mathbf{G},\,\sigma\sigma^\prime}$ are the `hopping' matrix elements between `sites' in momentum space (allowed states $\ket{\mathbf{q},\sigma}$). For a set of matrix elements which give a net complex phase upon hopping round a closed set of transfers in momentum space, time-reversal symmetry (TRS) is broken, and the resulting OFL can be topologically non-trivial. This is analogous to the Haldane model \cite{haldane_model_1988}, except that here it is hoppings in momentum space, rather than real-space, that are responsible for breaking TRS. It is important to note that TRS remains broken even for non-vanishing kinetic energy; the strong-coupling limit is not required to realise a topological bandstructure. In fact, we will more often work in the opposite weak-coupling limit in the remainder of this work.

We note again that while OFLs were first developed in the context of cold atoms, the same Hamiltonians arise in the study of twisted Moiré materials \cite{Wu2019}, although with less experimental flexibility in the choice of parameters of the potential \cite{Sommer_Ideal_2025}.

\subsection{Reciprocal-Space Construction}

\begin{figure}[t]
    \centering
    \sidesubfloat[]{\includegraphics[width=0.85\columnwidth]{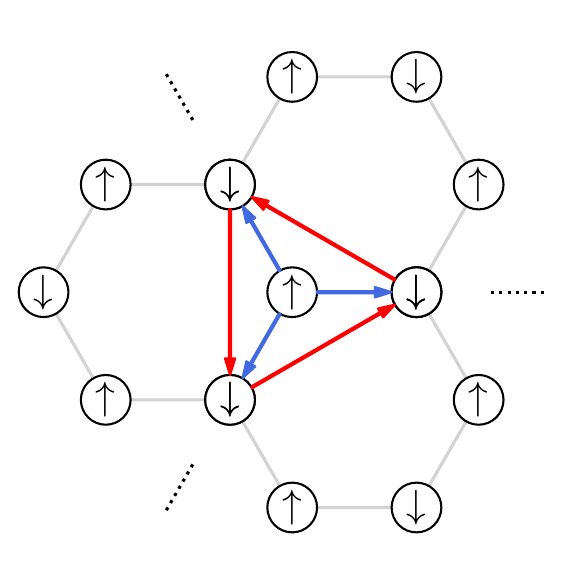}\label{Fig:Haldane_Couplings}}\hfill
    \sidesubfloat[]{\includegraphics[width=0.85\columnwidth]{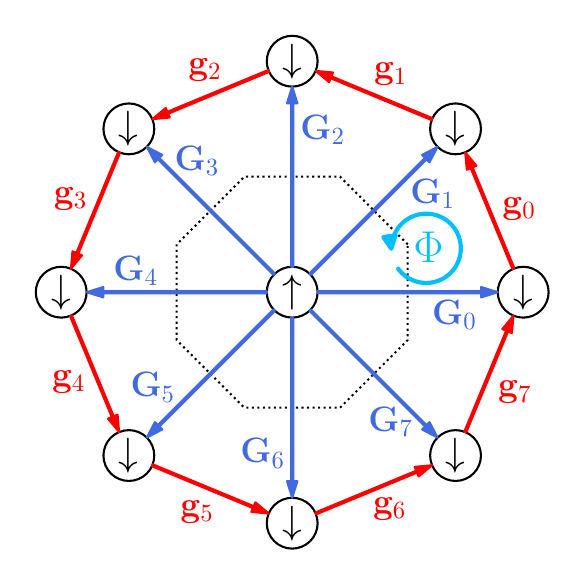}\label{Fig:Plaquette_Fluxes}}\hfill
    \caption{Couplings in momentum space for (a) the ``dual Haldane model'' and (b) our quasiperiodic model. Upon hopping anticlockwise around a triangular plaquette, the atom acquires a phase $\Phi$ from the complex coupling elements. The quasi-Brillouin zone (see Sec. \ref{Sec:PlaneWave}) is shown dotted in black.}
\end{figure}

We now construct the specific model that we will consider for the remainder of the paper. This is an adaptation of the ``dual Haldane model'' presented in \cite{Cooper_Optical_2011a}, in which the couplings of the potential in momentum space [Fig. \ref{Fig:Haldane_Couplings}] take the form of the Haldane model, with $\ket{\sigma=\,\uparrow}$ and $\ket{\sigma=\,\downarrow}$ states occupying the two inequivalent sites of the honeycomb lattice. There, each $\ket{\uparrow}$ site is coupled to three nearest-neighbour $\ket{\downarrow}$ sites with 3-fold rotational symmetry, in addition to next-nearest-neighbour coupling between like spins. Here, we make a quasiperiodic potential by coupling the $\ket{\uparrow}$ site to eight neighbouring $\ket{\downarrow}$ sites with 8-fold rotational symmetry, as well as retaining the couplings between like spins [Fig. \ref{Fig:Plaquette_Fluxes}]. Our Hamiltonian takes the form:
\begin{subequations}\label{Eq:Hamiltonian}
    \begin{equation}
        H = \frac{\mathbf{p}^2}{2m}\otimes\mathbbm{1}_2 + H_U + H_V
    \end{equation}
    \begin{equation}\label{Eq:H_U}
        H_U = \sum_{l=0}^{7} 
                U_l \, \mathrm{e}^{-\mathrm{i} \mathbf{G}_l \cdot \mathbf{r}} 
                \otimes \sigma_+ 
                + \text{H.c.}
    \end{equation}
    \begin{equation}\label{Eq:H_V}
        H_V = \sum_{l=0}^{7} 
                V_l \, \mathrm{e}^{-\mathrm{i} \mathbf{g}_l \cdot \mathbf{r}} 
                \otimes \sigma_z 
                + \text{H.c.}
    \end{equation}
\end{subequations}
where $\sigma_+=(\sigma_x+\textrm{i}\sigma_y)/2$ and $\sigma_{x,y,z}$ are the  Pauli matrices. $H_U$ generates the coupling between opposite spins, while $H_V$ couples like spins. The Fourier components of the potential have the desired 8-fold rotational symmetry:
\begin{subequations}
    \begin{equation}
        \mathbf{G}_l=|\mathbf{G}|\,(\cos(\pi l/4),\,\sin(\pi l/4))^\text{T}
    \end{equation}
    \begin{equation}
        \mathbf{g}_l=\mathbf{G}_{l+1}-\mathbf{G}_{l}
    \end{equation}
\end{subequations}
with $l\in [0,7]$. (We have also analysed an analogous system with 5-fold rotational symmetry, which shows qualitatively similar results; see Appendix \ref{App:Pentagonal}.) The wavelength $\lambda=2\pi/|\mathbf{G}|$ provides the natural lengthscale of the system. Since these $\mathbf{G}_l$ are incommensurate with one another, the minimal rank of the basis set spanning the potential (4) exceeds the spatial dimension (2), and the system is quasicrystalline by definition \cite{Lifshitz_Quasicrystals_2003}. Note that since we also include the kinetic energy term, our Hamiltonian is not simply a tight-binding model in momentum space, but rather a continuum model of the form of Eq. (\ref{Eq:GeneralHamiltonian}).

Analogous to the Haldane model, we wish to choose couplings which break TRS due to the non-zero complex phase $\Phi$ acquired upon hopping round any closed set of momentum transfers. For quasicrystals, there is an additional complication that combinations of basis vectors $\mathbf{G}=\sum_l n_l \mathbf{G}_l$ with $n_l \in \mathbb{Z}$ fill reciprocal space densely, instead of forming the sparse reciprocal lattice of a periodic system as in Fig. \ref{Fig:Haldane_Couplings}. For simplicity, we only show the `first-order' couplings in Fig. \ref{Fig:Plaquette_Fluxes}; nevertheless, these are sufficient to enable us to choose $U_l$ and $V_l$ elements for which TRS is broken. 

We choose constant $V_l=V\:\forall\,l$ for simplicity. Since $\mathbf{g}_l=-\mathbf{g}_{l+4}$, the effective matrix element for a transfer by momentum $\mathbf{g}_l$ is:
\begin{equation}
\begin{aligned}
    \bra{\mathbf{q}+\mathbf{g}_l, \, \sigma^\prime} H_V \ket{\mathbf{q}, \, \sigma} 
    &= (V_l^* + V_{l+4}) \bra{\sigma^\prime}\sigma_z \ket{\sigma}  \\
    &= 2\,\text{Re}[V] \bra{\sigma^\prime}\sigma_z \ket{\sigma},
\end{aligned}
\end{equation}
so we may pick $V\!\mathrel{\in}\!\mathbb{R}$ without loss of generality. To retain the rotational symmetry and ensure that the phase is the same for each triangular plaquette in Fig. \ref{Fig:Plaquette_Fluxes}, we choose: 
\begin{equation}
    U_l = -U\textrm{e}^{-\textrm{i} \pi R l/4}
\end{equation}
with $R\in\mathbb{Z}$. When starting from a $\ket{\uparrow}$ state, this results in a total acquired phase:
\begin{equation}
    \Phi_\uparrow=\text{arg}(U_l^*\cdot
    (-2V)
    \cdot U_{l+1})=\text{mod}\left(-\frac{\pi R}{4},\,2\pi\right)-\pi
\end{equation}
with the modulus operation ensuring that $-\pi\leq\Phi_\uparrow<\pi$. Alternatively, we could consider starting from a $\ket{\downarrow}$ state (not depicted in Fig. \ref{Fig:Plaquette_Fluxes}), in which case the acquired phase would be:
\begin{equation}
    \Phi_\downarrow=\text{arg}(U_l\cdot2V\cdot U_{l+1}^*)=\text{mod}\left(\frac{\pi R}{4}+\pi,\,2\pi\right)-\pi
\end{equation}
Throughout this work, we take $R=5$, giving $\Phi_\uparrow=-\pi/4$ and $\Phi_\downarrow=-3\pi/4$; we will find that this leads to a Chern number $\mathcal{C}=+1$. We emphasise that it is really $H_V$ which breaks TRS due to its $\sigma_z$ coupling; without this, the phases from $H_U$ would cancel out as $\Phi_\uparrow=-\Phi_\downarrow$. The associated magnetic field $B_{\text{eff}}(\mathbf{r})$ of the model inherits the quasiperiodic octagonal symmetry of the Hamiltonian~\footnote{Numerical evaluation shows that the spatial average of the magnetic field is $\langle B_{\text{eff}}\rangle_\mathbf{r}\approx(1.14\pm0.12)\times2\pi/\lambda^2$ (estimated by sampling different patches of the quasicrystal). However, for the periodic approximants of Sec.~\ref{Sec:Approximant}, the net flux per unit cell is undefined due to points of degeneracy between spinor eigenstates.}.

%% file: Experimental.tex
\section{Implementation for Cold Atoms}\label{Sec:Experimental}

\begin{figure*}[ht]
    \centering
    \sidesubfloat[]{\includegraphics[width=0.51\textwidth]{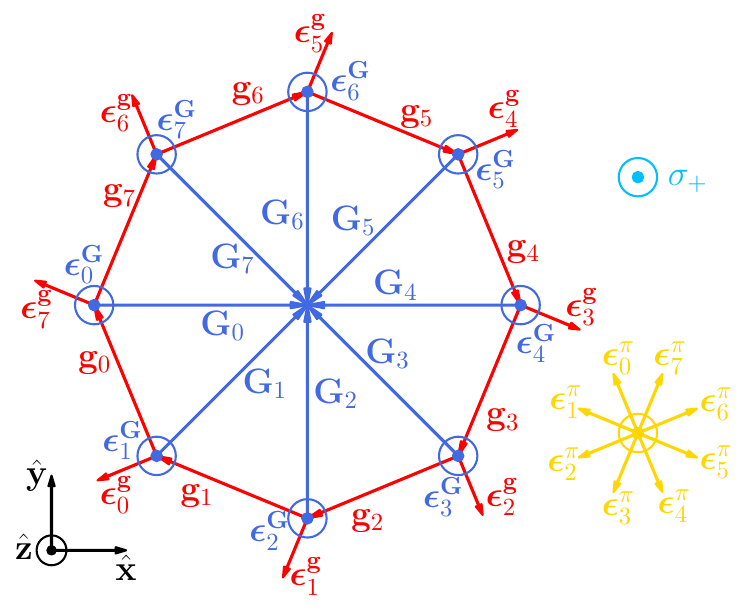}\label{Fig:Laser_Geometry}}\hfill
    \sidesubfloat[]{\includegraphics[width=0.38\textwidth]{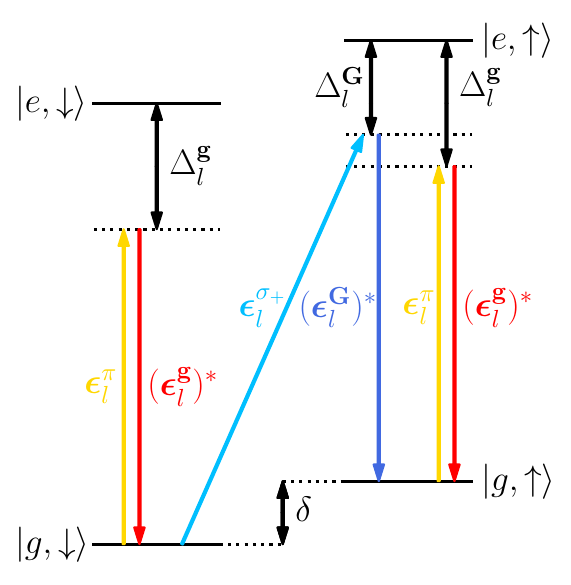}\label{Fig:Laser_Transitions}}
    \caption{(a) Laser wavevectors and electric field directions. The $\boldsymbol{\epsilon}_l^{\sigma +}$ (light blue) and $\boldsymbol{\epsilon}_l^{\pi}$ (yellow) lasers propagate in the $z$-direction, out of the plane. The circularly-polarised $\boldsymbol{\epsilon}_l^{\sigma +}$ electric fields (not depicted) rotate in the $xy$-plane; all other lasers are linearly polarised along the directions indicated. (b) Illustration of the energy levels and transitions contributing to the Hamiltonian. The coloured arrows indicate absorption or emission of a photon from the corresponding laser in (a). The $\boldsymbol{\epsilon}_l^{\sigma +}$/$\boldsymbol{\epsilon}_l^{\mathbf{G}}$ (light/dark blue) process generates the $H_{U}$ terms, and the $\boldsymbol{\epsilon}_l^{\pi}$/$\boldsymbol{\epsilon}_l^{\mathbf{g}}$ (yellow/red) process generates the $H_{V}$ terms.}
    \label{Fig:Model_System}
\end{figure*}

We interrupt the analysis of our model to detail how it may be realised in a cold-atom setting. This section is self-contained, and not required for understanding the model; readers not interested in this experimental implementation may skip to Sec. \ref{Sec:PlaneWave}. We emphasise again that other experimental realisations, for example in twisted Moiré materials, may be possible.

Our strategy will be to use two-photon Raman transitions to generate the required momentum shifts and couplings between spin states, which in general may be any two angular momentum states. (In the 2D materials setting~\cite{Wu2019}, the role of spin is provided by the layer index.) For simplicity of presentation, we consider an atomic species with a ground level with angular momentum $1/2$, with states denoted $\{|g,\uparrow\rangle,|g,\downarrow\rangle\}$, and an excited level also of angular momentum $1/2$, with states $\{|e,\uparrow\rangle,|e,\downarrow\rangle\}$. Possible atomic species include $^{171}\text{Yb}$ or $^{199}\text{Hg}$ \cite{Cooper_Optical_2011a}. We imagine that the spin-degeneracy is lifted, for example due to the Zeeman shift from an applied magnetic field, resulting in splitting of both the ground and excited levels by an energy $\delta$. 

We characterise our lasers by complex electric field amplitudes $\boldsymbol{\epsilon}\in\mathbb{C}^3$, such that the physical electric field is $\mathbf{E}(\mathbf{r},t)=\text{Re}[\boldsymbol{\epsilon}\,\textrm{e}^{\textrm{i}(\mathbf{k}\cdot\mathbf{r} - \omega t)}]$. For a two-photon process involving absorption from laser $i$ and emission into laser $j$, the `vector light shift' gives a  potential energy term:
\begin{equation}\label{Eq:Laser_Potential}
    V(\mathbf{r}) \propto \left(\frac{\boldsymbol{\epsilon}_i^* \times \boldsymbol{\epsilon}_j} {\Delta} \cdot\boldsymbol{\sigma} \right)
    \textrm{e}^{\textrm{i}\Delta \mathbf{k}\cdot\mathbf{r}}
\end{equation}
where $\Delta \mathbf{k}=\mathbf{k}_i-\mathbf{k}_j$ is the momentum kick acquired from the two photons, $\boldsymbol{\sigma} = (\sigma_x,\sigma_y,\sigma_z)^{\text{T}}$ is the vector of Pauli spin matrices, and $\Delta$ is the detuning of the transition energy from the excited state (taken to be large enough that any population of the excited state can be neglected). We refer the reader to Refs.  \cite{deutschQuantumstateControlOptical1998, Cooper_Topological_2019} for the derivation of Eq. (\ref{Eq:Laser_Potential}). We build up our desired potential [Eq. (\ref{Eq:Hamiltonian})] term-by-term by adding pairs of lasers to our setup. 

The full diagram of lasers and the corresponding atomic energy transitions are shown in Fig. \ref{Fig:Model_System}. To generate the $H_U$ terms, we couple lasers with wavevectors $\mathbf{G}_l$ and $\boldsymbol{\epsilon}_l^\mathbf{G}$ linearly polarised along $z$, to lasers propagating along $z$ with $\boldsymbol{\epsilon}_l^{\sigma_+}$ circularly polarised in the $xy$-plane. We take the $z$ momentum kick to be suppressed by an external trapping potential, so the net momentum transfer takes the desired form $\textrm{e}^{-\textrm{i}\,\mathbf{G}_l\cdot\mathbf{r}}$; likewise, the choices of electric field polarisations pick out the desired $\sigma_+$ coupling in spin-space. The associated amplitude is:
\begin{equation}
     U_l \propto \frac{|(\boldsymbol{\epsilon}_l^{\mathbf{G}})^*\times \boldsymbol{\epsilon}_l^{\sigma_+}|}{\Delta_l^{\mathbf{G}}} \,.
\end{equation}
Similarly for the $H_V$ terms, we couple between in-plane lasers with wavevectors $\mathbf{g}_l$, and out-of-plane $z$-propagating lasers. We pick the respective electric fields $\boldsymbol{\epsilon}_l^\mathbf{g}$ and $\boldsymbol{\epsilon}_l^\pi$ both to be linearly polarised in the $xy$-plane, but orthogonal to each other. This gives the required momentum components $\textrm{e}^{-\textrm{i}\,\mathbf{g}_l\cdot\mathbf{r}}$ and $\sigma_z$ spin coupling, with amplitudes:
\begin{equation}
     V_l \propto \frac{|(\boldsymbol{\epsilon}_l^{\mathbf{g}})^*\times \boldsymbol{\epsilon}_l^{\pi}|}{\Delta_l^{\mathbf{g}}} \,.
\end{equation}
This scheme allows us to choose any $U_l, \, V_l\in\mathbb{C}$ by adjusting the laser amplitudes, detunings and relative phases. 

We imagine choosing different detunings $\Delta_l^{\mathbf{G}/\mathbf{g}}$, so all lasers are incoherent except for the 16 pairs generating the matrix elements $U_l$ and $V_l$. This realises exactly our Hamiltonian Eq. (\ref{Eq:Hamiltonian}), but may be challenging experimentally. An alternative scheme is outlined in Appendix \ref{App:Coherent}, which requires only five lasers, all of which are coherent. This comes at the cost of making $U_l$ and $V_l$ dependent on one another, and introducing additional momentum couplings into the Hamiltonian. We have analysed both cases and find similar qualitative behaviour, but we will focus on the first scheme for maximum clarity.

%% file: PlaneWave.tex
\section{Plane-Wave Analysis}\label{Sec:PlaneWave}

\subsection{Quasi-Brillouin Zone}\label{Sec:QBZ_Definition}

We now analyse our model to demonstrate that it does possess the desired non-trivial topological properties. To provide analytic insight, we first consider the system in the `nearly-free' limit of weak potential applied in Refs. \cite{Spurrier_Semiclassical_2018, Gambaudo_Brillouin_2014}. We shall later show that the (topological) features derived within this analytic approach survive also to deeper lattice depths. 

For $U\!=\!V\!=\!0$, the energy eigenstates of the system are simply the plane-wave states $\ket{\mathbf{q},\sigma}$, with $\sigma \in \{\uparrow, \downarrow\}$ as before. The energy levels form the (spin-degenerate) parabola $E(\mathbf{q})=|\mathbf{q}|^2/2m$. We first introduce $H_U$ only, with a small $U\ll E_R$, where $E_R=|\mathbf{G}|^2/2m$ is the recoil energy which sets the kinetic energy scale of the system. This couples each state $\ket{\mathbf{q},\sigma}$ to other states $\ket{\mathbf{q}-\mathbf{G}_l,\sigma^\prime}$ (with $\sigma^\prime$ opposite to $\sigma$). This opens gaps in the parabola at first-order in $U$ when the states have equal kinetic energy. Such states lie along the perpendicular bisectors of the $\mathbf{G}_l$ vectors. In a periodic system where $\mathbf{G}_l$ are reciprocal lattice vectors, these gaps define the boundary of the Brillouin zone; here, gaps define an octagonal region of reciprocal space, depicted in Fig. \ref{Fig:Plaquette_Fluxes}, that we term the `quasi-Brillouin zone' (QBZ) as it  does not tile the plane. The low-energy states with momenta inside the QBZ therefore form a band separated from the rest of the spectrum which, as we show below, can be made topological.

Just as the Brillouin zone area (in reciprocal space) determines the number of states in a band in a periodic system, the QBZ area determines the number of states in our band here. Imposing periodic boundary conditions over an arbitrary area $L^2$, allowed states occupy an area $(2\pi/L)^2$ in reciprocal space, so the required atomic density to fill all the states is calculated geometrically to be:
\begin{equation}\label{Eq:Density}
    n=\frac{2A_\text{QBZ}}{(2\pi)^2}=\frac{4(\sqrt{2}-1)}{\lambda^2}\approx \frac{1.6568\ldots}{\lambda^2}
\end{equation}
with the extra factor of 2 due to spin. We further verify this result in Sec. \ref{Sec:Density}. 

We pause to note that this momentum space approach gives a useful alternative perspective on the model of Ref. \cite{Gottlob_Hubbard_2023}. That work studies a similar 8-fold rotationally symmetric quasicrystal with a real-space methodology, numerically finding a (non-topological) isolated lowest band. With our methodology, we can physically understand the emergence of this isolated band due to the opening of gaps along the QBZ boundary of that system. After accounting for spin, and inserting the appropriate units of lengths, our expression for the atomic density to fill the band [Eq. (\ref{Eq:Density})] agrees with that calculated by an alternative method in Ref. \cite{Gottlob_Hubbard_2023}.

\subsection{Symmetry and Topological Properties}\label{Sec:QBZ_Bandstructure}

\begin{figure}[t]
    \centering
    \sidesubfloat[]{\includegraphics[width=0.85\columnwidth]{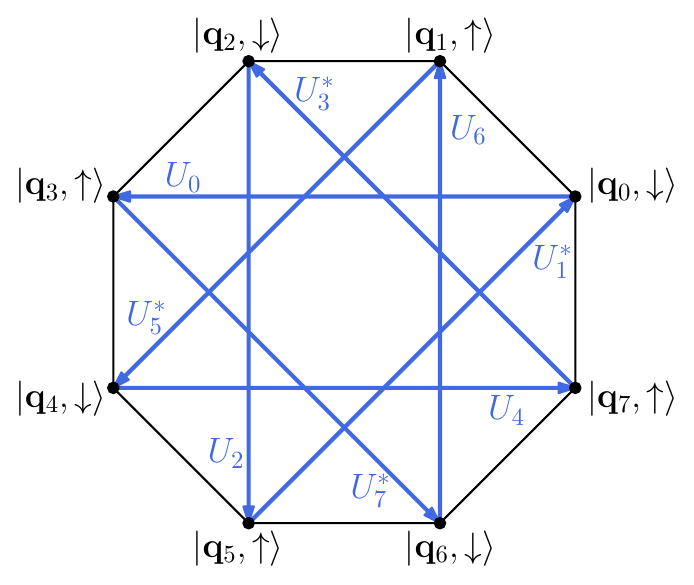}\label{Fig:QBZ_Couplings_U}}\hfill
    \sidesubfloat[]{\includegraphics[width=0.85\columnwidth]{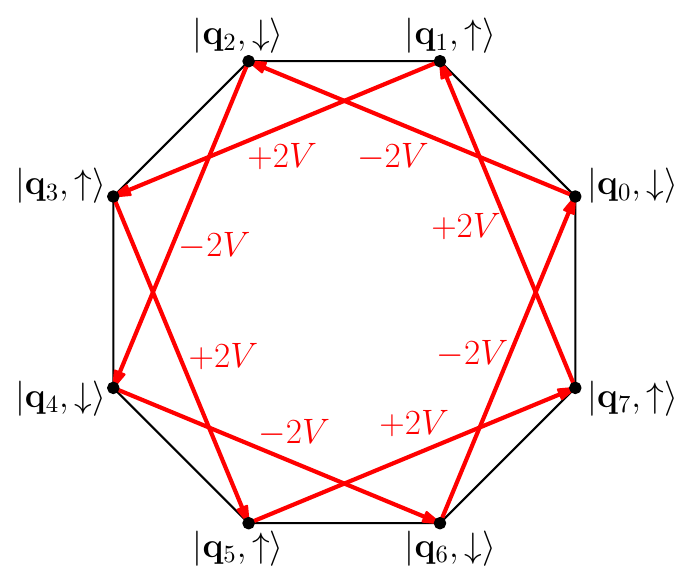}\label{Fig:QBZ_Couplings_V}}\hfill
    \caption{Couplings of corners in the QBZ. The $H_U$ and $H_V$couplings are shown on different diagrams (a) and (b) for clarity. The set of 8 states with opposite spin to those indicated here are coupled by different $U_l$ and $V_l$ terms; there is no coupling between the two sets.}
    \label{Fig:QBZ_Couplings}
\end{figure}

\begin{figure}[t]
    \centering
    \sidesubfloat[]{\includegraphics[width=0.85\columnwidth]{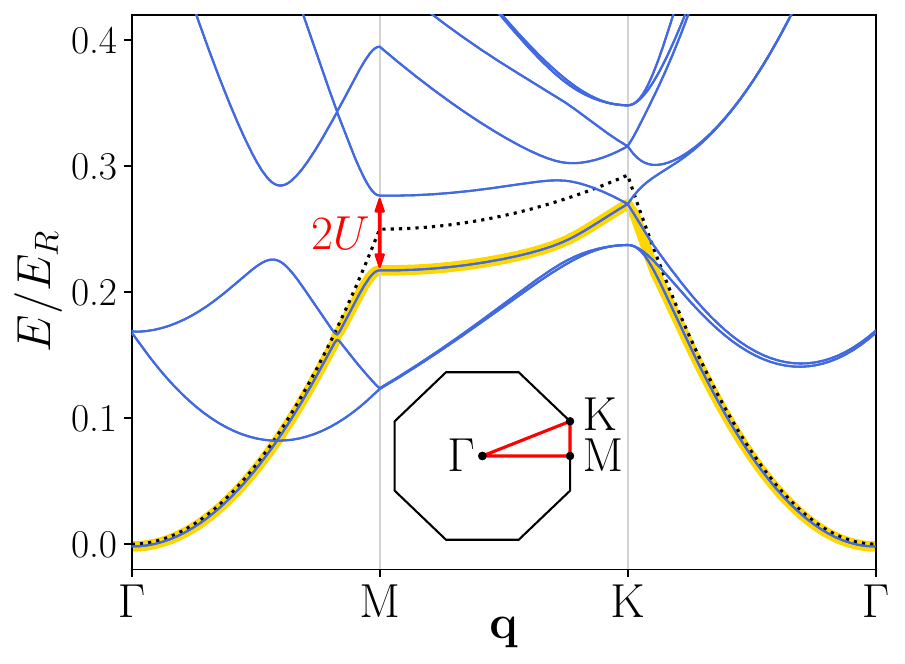}\label{Fig:HU_Bandstructure}}\hfill
    \sidesubfloat[]{\includegraphics[width=0.85\columnwidth]{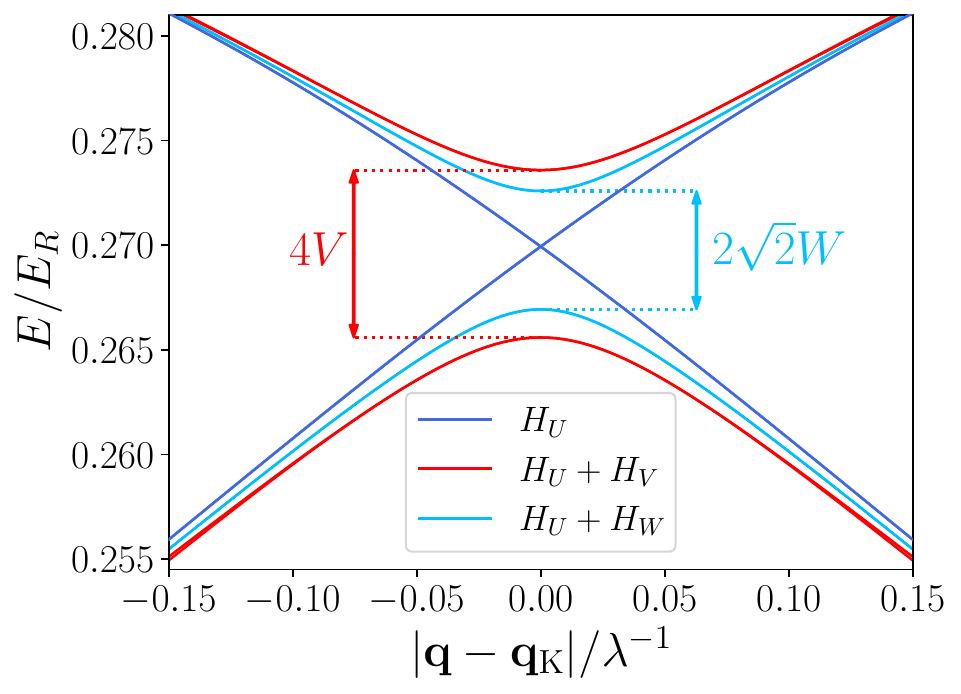}\label{Fig:K_Bandstructure}}\hfill
    \caption{(a) Bandstructure across the QBZ, calculated in the basis of the 16 plane-wave states degenerate at the $\text{K}$ point, for $U/E_R=0.03,\,V=W=0$. The path in the QBZ is shown inset. All bands here are doubly degenerate. The relevant states are those highlighted in yellow which tend to the free-particle dispersion (black dotted curve) as $U\to0$. (b) Bandstructure in the vicinity of the $\text{K}$ point, along the line $\Gamma \text{K}$. The bands are calculated for $H_U$ only (dark blue), $H_U+H_V$ with $V/E_R=0.002$ (red), and $H_U+H_W$ with $W/E_R=0.002$ (light blue).}
    \label{Fig:QBZ_Bandstructure}
\end{figure}

Having identified this isolated set of states, it is natural to investigate whether the band that they form can be topologically non-trivial. To do so, we compute the bandstructure over the QBZ in the minimal basis of plane-wave states which captures all first-order couplings of the Hamiltonian. This requires 16 basis states, which are degenerate and coupled together at the QBZ corners. Explicitly, the required states $\ket{\mathbf{q}_i,\sigma}$ are related by $\mathbf{q}_{i+3\,\text{(mod 8)}}=\mathbf{q}_i-\mathbf{G}_i$. The couplings are illustrated in Fig. \ref{Fig:QBZ_Couplings}. Diagonalising the $16\times 16$ Hamiltonian, we find the bandstructure in Fig. \ref{Fig:QBZ_Bandstructure}. The relevant bands to focus on are those (highlighted in yellow in Fig. \ref{Fig:HU_Bandstructure}) that tend to the free-particle parabolic dispersion in the limit of $U\!\to\! 0$. These are where the spectral density (see Sec. \ref{Sec:Spectral}) is concentrated in this weak-coupling regime, since the bare free-particle states $\ket{\mathbf{q},\sigma}$ remain good eigenstates away from points of degeneracy with other basis states (that is, away from the QBZ boundary).

As expected, along the QBZ boundary, we generally find a gap between the relevant bands. The exception is at the corners $\text{K}$, where the symmetry of $H_U$ under various operations involving spatial rotations and reflections combined with spin transformations enforces a four-fold degeneracy of states. We direct the reader to Appendix \ref{App:Symmetries} for a detailed presentation of this symmetry analysis. In the vicinity of the corners [Fig. \ref{Fig:K_Bandstructure}], the dispersion is linear; in other words, we have two coincident Dirac cones in the bandstructure. Thus, the potential $H_U$ leads to a semimetallic quasicrystalline system, with a density of states that vanishes linearly in energy, at least within this weak potential limit. 

This semimetallic quasicrystalline state is analogous to the graphene bandstructure on which the Haldane model is built. (We summarise the relevant details of the Haldane model in Appendix \ref{App:Haldane_Symmetries}.) Like in that system, we can consider adding terms to our Hamiltonian which break symmetries and open a gap at the Dirac point of this quasicrystalline lattice. A gapped Dirac cone is generically associated with a Berry flux of $\pm \pi$; the topological character of the states below the gap therefore depends on whether we have the same or opposite sign of flux for each Dirac cone, corresponding to Chern number $|\mathcal{C}|=1$ or 0 respectively. This in turn depends on whether the perturbation breaks or preserves TRS.

One possible perturbation is $H_V$ [Eq. (\ref{Eq:H_V})], which breaks TRS by design (analogous to the next-nearest-neighbour hopping in the Haldane model). For our choice $R=5$ and $V>0$, both bands below the gap contribute a Berry flux of $+\pi$, resulting in $\mathcal{C}=1$. This demonstrates that our model Hamiltonian Eq. (\ref{Eq:Hamiltonian}) does indeed describe a topological quasicrystal. To illustrate that TRS-preserving perturbations (analogous to the Semenoff mass) are also possible, we consider:
\begin{equation}\label{Eq:H_W}
    H_W = W\sum_{l=0}^{7} (-1)^{l} \,  \textrm{e}^{-\textrm{i}\mathbf{G}_l\cdot\mathbf{r}} \otimes\mathbbm{1}_2 + \text{H.c.}
\end{equation}
with $W\in\mathbb{R}$. This introduces new couplings between basis states (breaking the symmetry $S$ defined in Appendix \ref{App:Symmetries}) and gaps the Dirac points, but does not break TRS, so the bands below the gap have opposite signs of Berry curvature and the system is topologically trivial. See Appendix \ref{App:Symmetries} for further details of the exact symmetries that are broken by $H_V$ and $H_W$ to gap out the Dirac points, and comparisons to their counterparts in the Haldane model. When projecting to the degenerate subspace at the Dirac point, we find an effective coupling strength of $2V$ for $H_V$, and $\sqrt{2}W$ for $H_W$, giving rise to the different gaps in Fig. \ref{Fig:K_Bandstructure}. In general, when both $V$ and $W$ are non-zero, the Chern number is simply determined by whichever term dominates:
\begin{equation}
    \mathcal{C}=
    \begin{cases}
    1 & \sqrt{2}V>W \,,\\
    0 & W>\sqrt{2}V\,.
    \end{cases}
\end{equation}
It is important to note that since $\mathcal{C}$ is a topological invariant, it cannot change as long as the relevant energy gap does not close. As such, although we have used the weak-coupling limit to derive the value $\mathcal{C}=1$, the topological behaviour can be expected to persist as $U$ and $V$ increase to stronger coupling strengths.

\subsection{Extended Plane-Wave Basis}\label{Sec:Spectral}
We now calculate the energy spectrum of the quasicrystal using an extended basis of plane waves $\{\ket{\mathbf{q}-\mathbf{b}_i,\,\sigma}\}$, where:
\begin{equation}
    \mathbf{b}_i = \sum_ln_{il}\mathbf{G}_l
\end{equation}
with $n_{il}\in\mathbb{Z}$. We truncate the basis by imposing a maximum $N=\sum_l |n_{il}|$, ensuring that our calculations are converged with respect to this cutoff. This captures higher-order couplings of the Hamiltonian, allowing us to investigate the range of validity of the weak-coupling (first-order) approximation made in the previous section.

\begin{figure}[t]
    \centering
    \sidesubfloat[]{\includegraphics[width=0.85\columnwidth]{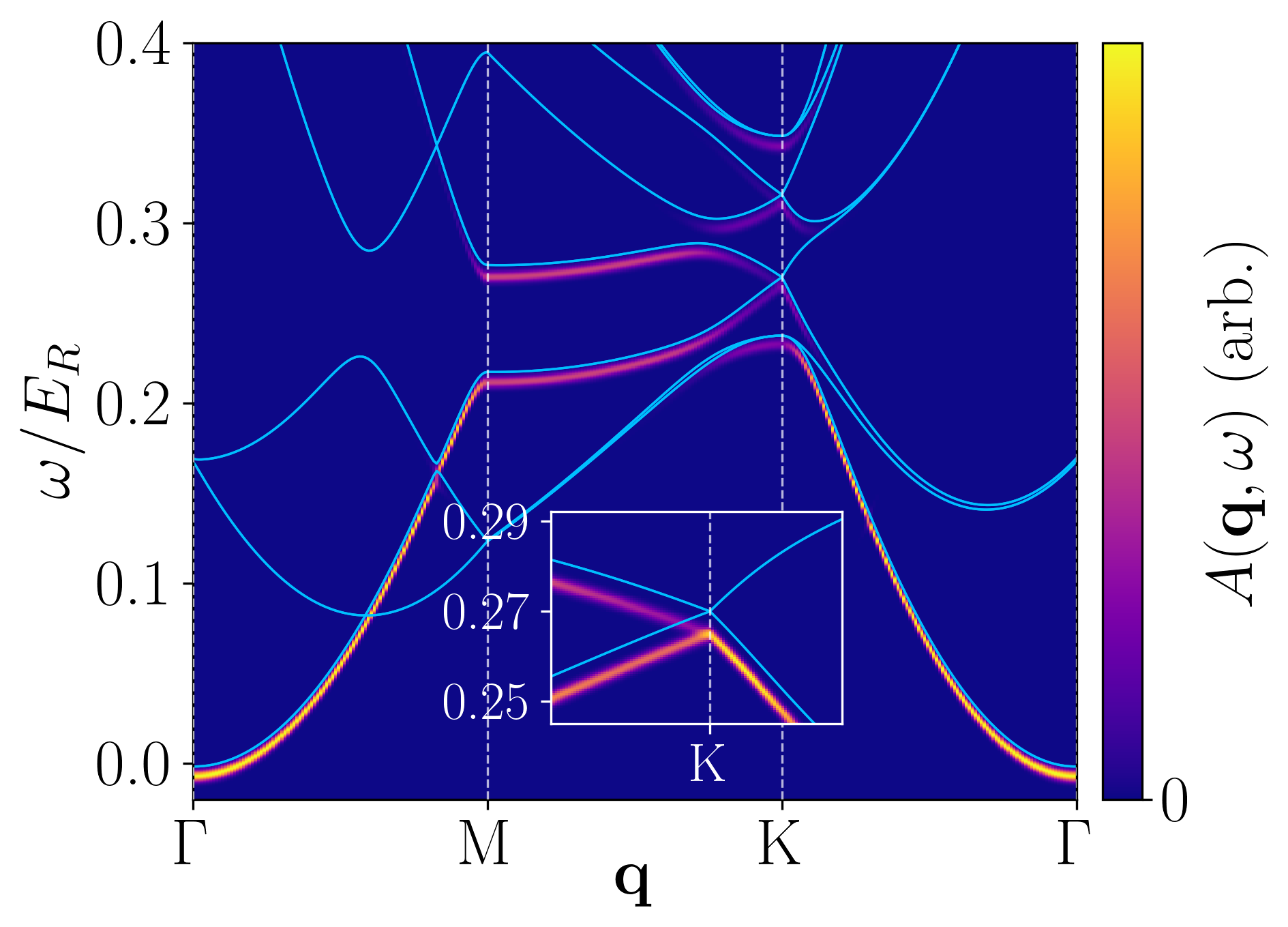}\label{Fig:Spectral_U0.03}}\hfill
    \sidesubfloat[]{\includegraphics[width=0.85\columnwidth]{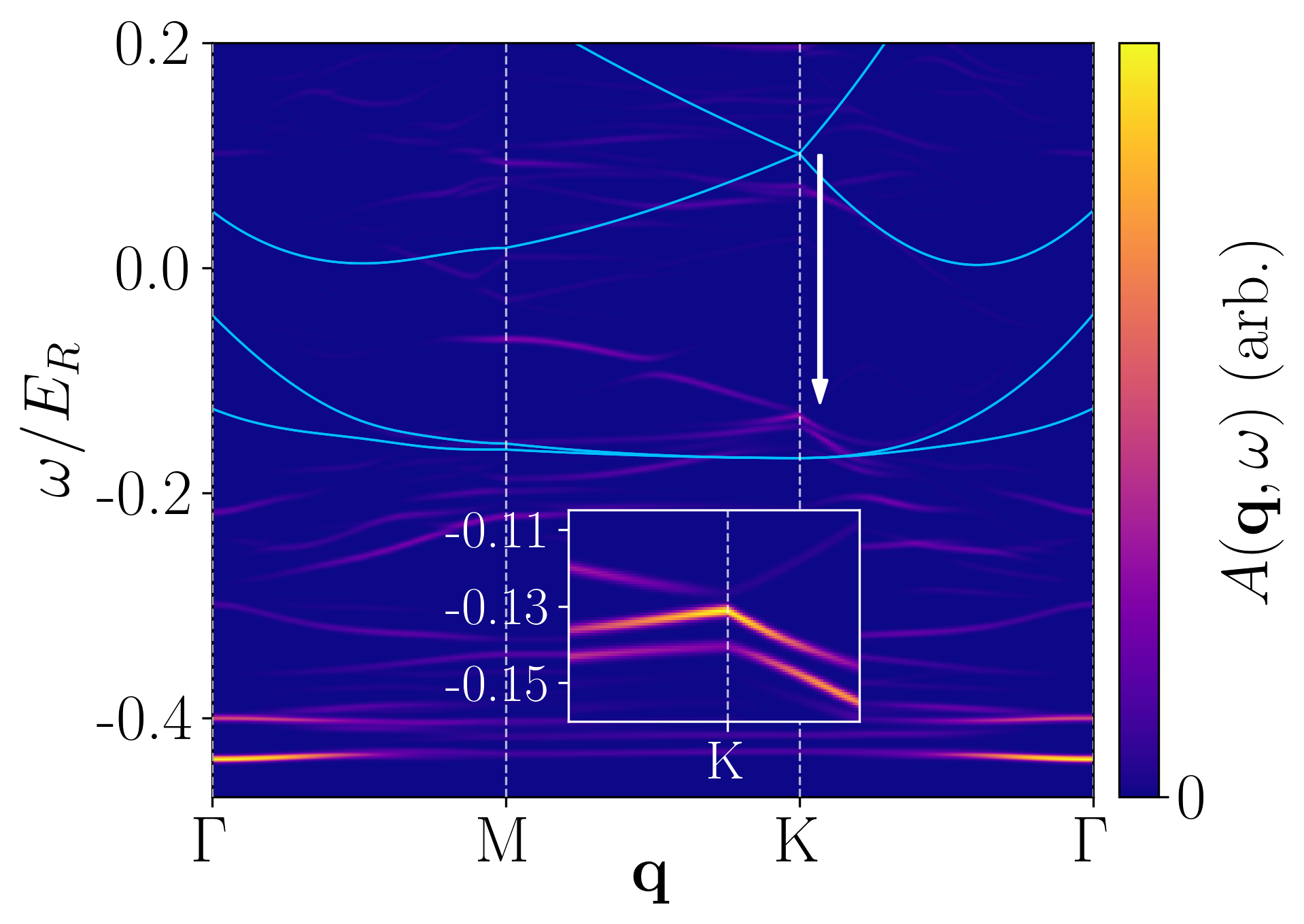}\label{Fig:Spectral_U0.25}}\hfill
    \caption{Spectral function calculated with the extended plane-wave basis, for (a) $U/E_R\!=\!0.03$ and (b) $U/E_R\!=\!0.25$; $V\!=\!W\!=\!0$ in both plots. The path in momentum space is the same as in Fig. \ref{Fig:HU_Bandstructure}. Blue lines show the bandstructure calculated for the same values of $U$ using the reduced basis of Sec. \ref{Sec:QBZ_Bandstructure}. This agrees well with $A(\mathbf{q},\omega)$ in (a); in (b), the Dirac point is visibly gapped, and its energy is shifted, as indicated by the white arrow. Insets show the vicinity of the K point; the colour scales of the insets are not the same as the main plots.}
    \label{Fig:Spectral_Function}
\end{figure}

Having diagonalised the Hamiltonian in this basis to obtain eigenstates $\ket{n}$ with eigenvalues $E_n$, we calculate the spectral function (summed over spin states):
\begin{equation}\label{Eq:Spectral_Function}
    A(\mathbf{q},\omega)=\sum_n \left[ \delta(\omega-E_n) \sum_{\sigma=\uparrow,\downarrow} |\!\braket{\mathbf{q},\sigma|n}\!|^2 \right].
\end{equation}
(In practice, we use a narrow Gaussian in place of the delta-function, resulting in non-zero line widths.) Two examples with different values of $U$ are plotted in Fig. \ref{Fig:Spectral_Function}, with $V\!=\!W\!=\!0$ in both cases. For $U/E_R\!=\!0.03$ [Fig. \ref{Fig:Spectral_U0.03}], within the weak-coupling limit, we find excellent agreement with the bandstructure calculated as in the previous section; in particular, we see the degenerate crossing of bands at the Dirac point. We find similarly good agreement in calculations with small but non-zero $V,W\ll E_R$ (not plotted here); the Dirac point gaps out as expected. 

For larger $U/E_R\!=\!0.25$ [Fig. \ref{Fig:Spectral_U0.25}], we see substantial departures from the reduced-basis bandstructure: the Dirac point is shifted in energy, and also becomes gapped. This is expected since the four-fold degeneracy is only symmetry-protected within the weak-coupling limit. However, it is interesting to note that although the energy of the Dirac point is substantially shifted, the four states remain near-degenerate, even at the moderate value of $U/E_R\!=\!0.25$. Therefore, in practice, we may expect topological behaviour to persist beyond the strict limit of $U\ll E_R$, with the same physical origin of Dirac points being gapped by a TRS-breaking perturbation.

%% file: Approximant.tex
\section{Approximant Analysis}\label{Sec:Approximant}

\subsection{Methodology}

\begin{figure}[t]
    \centering
    \includegraphics[width=0.9\columnwidth]{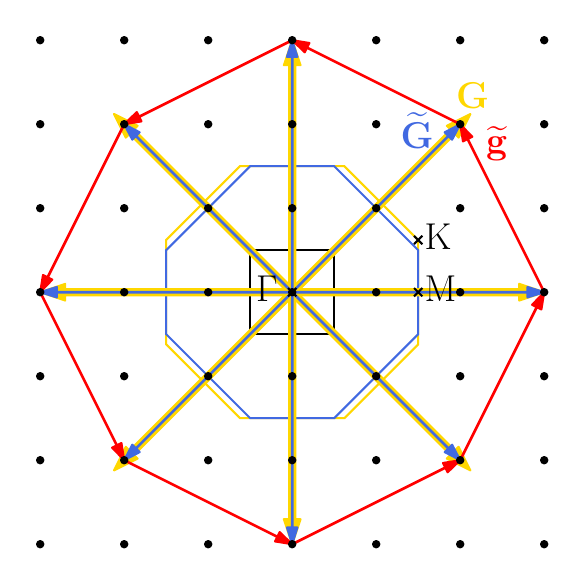}
    \caption{Illustration of the construction of the approximant vectors, for $N_{\rm a}=3$. The exact quasicrystal $\mathbf{G}_l$ are shown in yellow, and the approximant $\widetilde{\mathbf{G}}_l$ and $\widetilde{\mathbf{g}}_l$ in dark blue and red respectively. The approximant BZ is the black square centred on the origin; the approximant QBZ is the blue (irregular) octagon; and the quasicrystal QBZ is the yellow (regular) octagon. The high-symmetry points $\Gamma$, M, K of the quasicrystal QBZ are labelled with black crosses.}
    \label{Fig:Approximant_Diagram}
\end{figure}

To confirm our analytical plane-wave results and move beyond the weak-coupling limit, we numerically analyse approximant systems. These approximate the quasicrystalline geometry, but are in fact periodic over long lengthscales, and can therefore be investigated using standard techniques. We define a square lattice in reciprocal space with spacing $|\mathbf{G}|/N_{\rm a}$, with $N_{\rm a}\in\mathbb{N}$. The approximant $\mathbf{G}$-vectors $\widetilde{\mathbf{G}}_l$ are then found by `rounding' the quasicrystalline $\mathbf{G}_l$ to the nearest lattice point, thereby becoming commensurate with each other (see Fig. \ref{Fig:Approximant_Diagram}). Explicitly,
\begin{equation}
    \widetilde{\mathbf{G}}_l=\frac{\lfloor N_{\rm a} \mathbf{G}_l \rceil}{N_{\rm a}},
\end{equation}
where $\lfloor\mathbf{v}\rceil$ denotes rounding of each component of the vector $\mathbf{v}$ to the nearest integer. The approximant $\widetilde{\mathbf{g}}_l$ are given by  $\widetilde{\mathbf{G}}_{l+1} - \widetilde{\mathbf{G}}_l$ as before. The approximant Hamiltonian $\widetilde{H}$ has identical form to Eq. (\ref{Eq:Hamiltonian}), but with $\widetilde{\mathbf{G}}_l$ and $\widetilde{\mathbf{g}}_l$ substituted for $\mathbf{G}_l$ and $\mathbf{g}_l$. As all basis vectors are commensurate, $\widetilde{H}$ is periodic. The Brillouin zone (BZ) of the approximant model is the square region (in reciprocal space) $-\frac{|\mathbf{G}|}{2N_{\rm a}}<q_x,q_y<\frac{|\mathbf{G}|}{2N_{\rm a}}$, and the real space lattice constant is $a=N_{\rm a}\lambda$. The quasicrystal corresponds to the limit $N_{\rm a}\to\infty$ where the unit cell diverges. 

\subsection{Numerical Results}

We first calculate the spectral function [Eq. (\ref{Eq:Spectral_Function})] of the approximant to enable a direct comparison with the results of the previous section. The eigenspectrum is calculated by diagonalising the Bloch Hamiltonian $\widetilde{H}_\mathbf{k} = \textrm{e}^{\textrm{i}\mathbf{k}\cdot\mathbf{r}} \widetilde{H} \textrm{e}^{-\textrm{i}\mathbf{k}\cdot\mathbf{r}}$ using a plane-wave basis $\{\ket{\mathbf{k} + \widetilde{\mathbf{G}},\,\sigma}\}$, where $\{\widetilde{\mathbf{G}}\}$ are reciprocal lattice vectors of the approximant. We choose the basis to be large enough that our calculated quantities are converged. Here, $\mathbf{k}$ is the crystal momentum which lies inside the BZ; the calculation of $A(\mathbf{q},\omega)$ requires unfolding of energy bands over $\mathbf{k}$ to bands over the physical momentum $\mathbf{q}$.

\begin{figure}[t]
    \centering
    \sidesubfloat[]{\includegraphics[width=0.85\columnwidth]{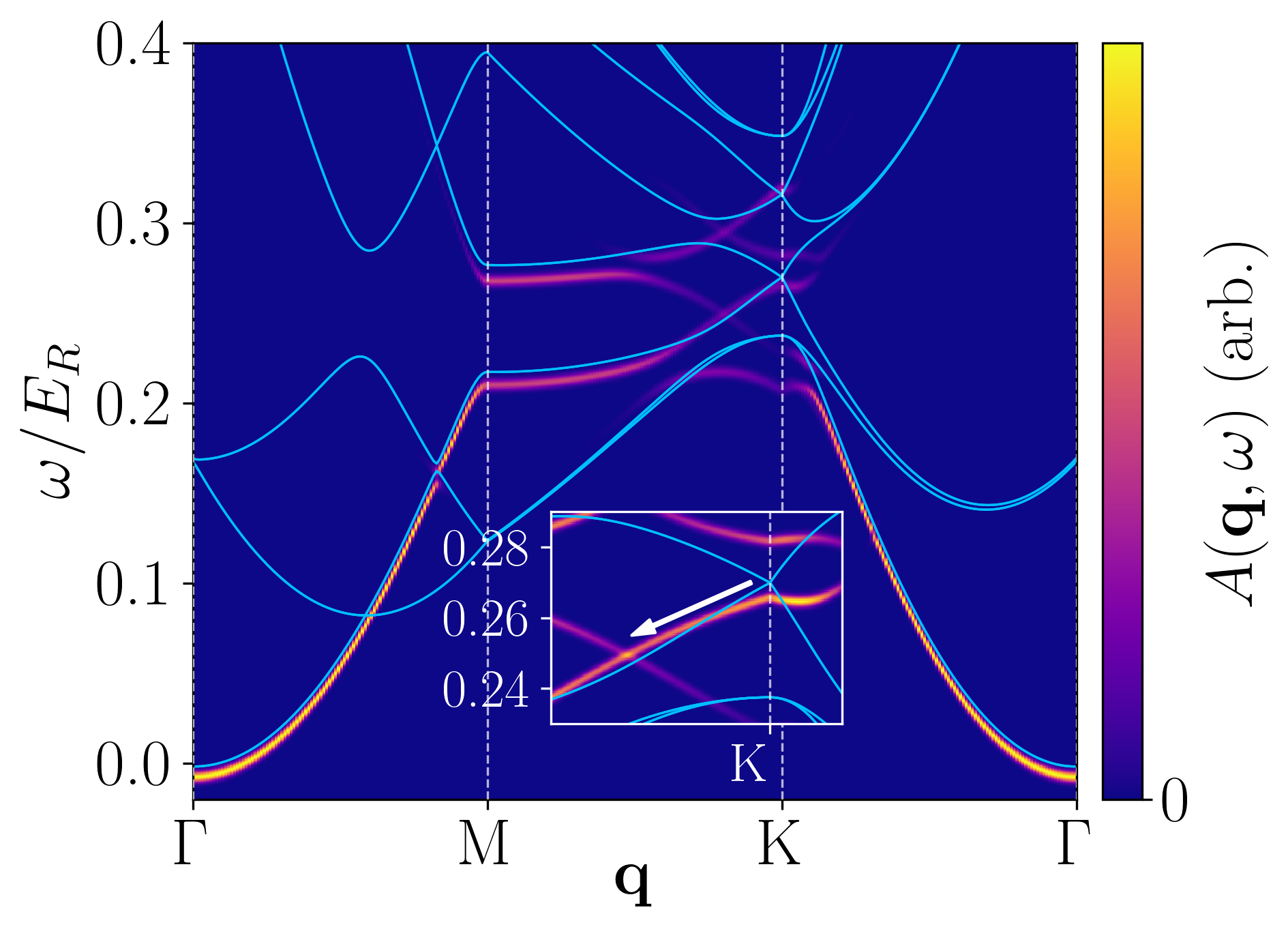}\label{Fig:Spectral_Approx_Na3}}\hfill
    \sidesubfloat[]{\includegraphics[width=0.85\columnwidth]{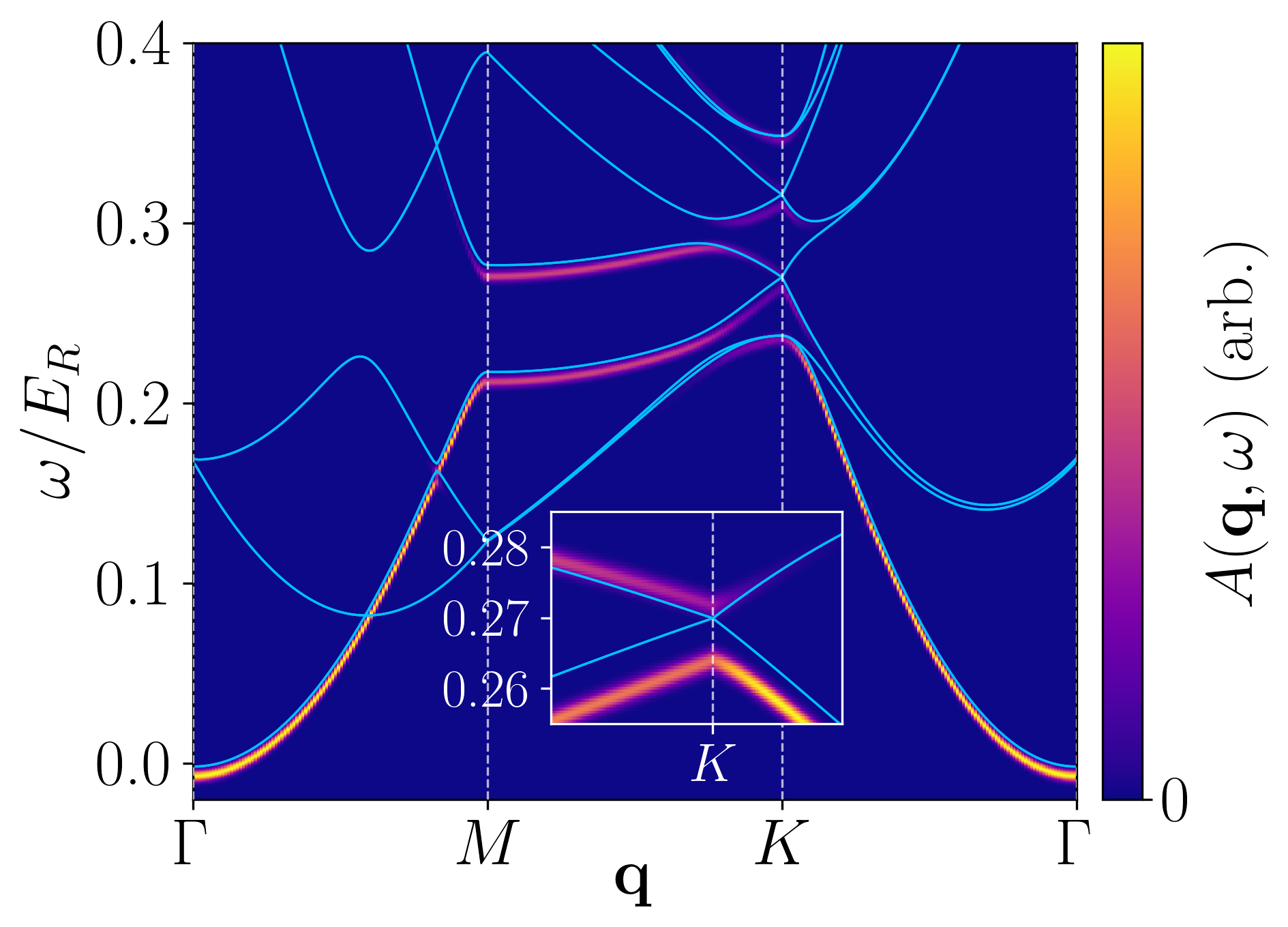}\label{Fig:Spectral_Approx_Na7}}\hfill
    \caption{Spectral function calculated by unfolding the bandstructure of approximant systems with (a) $N_{\rm a}\!=\!3$ and (b) $N_{\rm a}\!=\!7$. In both cases, $U/E_R\!=\!0.03$ and $V\!=\!W\!=\!0$. Blue lines show the bandstructure calculated in the reduced basis of plane waves as in Sec. \ref{Sec:QBZ_Bandstructure}. The path in momentum space is the same as in Fig. \ref{Fig:HU_Bandstructure}; that is, the points $\Gamma$, M and K refer to the quasicrystal QBZ. Insets show the vicinity of the K point; the colour scales of the insets are not the same as the main plots.}
    \label{Fig:Spectral_Function_Approximant}
\end{figure}

$A(\mathbf{q},\omega)$ is plotted in Fig. \ref{Fig:Spectral_Function_Approximant} for $N_{\rm a}\!=\!3$ and $N_{\rm a}\!=\!7$. For $N_{\rm a}\!=\!3$, the Dirac point is shifted not only in energy relative to the plane-wave bandstructure of Sec. \ref{Sec:QBZ_Bandstructure}, but also in momentum. This is because the Dirac point forms at the momentum where 16 basis states have degenerate kinetic energy; for the approximant with its $\widetilde{\mathbf{G}}_l$ vectors, this occurs at the corner of the `approximant QBZ' depicted in Fig. \ref{Fig:Approximant_Diagram} (and also discussed in Sec. \ref{Sec:Density}). As seen in Fig. \ref{Fig:Approximant_Diagram}, this point lies part-way along the MK line of the quasicrystal QBZ, explaining its position in Fig. \ref{Fig:Spectral_Approx_Na3}. For $N_{\rm a}\!=\!7$, the approximant $\widetilde{\mathbf{G}}_l$ vectors are closer to the exact quasicrystal $\mathbf{G}_l$ vectors, so the approximant Dirac point is closer to the quasicrystal K point. In fact, in this case the approximant Dirac point lies along the MK line of the quasicrystal but beyond the K point, so the apparent gap in the spectral function in Fig. \ref{Fig:Spectral_Approx_Na7} is much larger than the true gap at the Dirac point. 

For both approximants, the degeneracy at the Dirac point is not quite exact (although this appears to be the case with the energy resolution of Fig. \ref{Fig:Spectral_Approx_Na3}). This is because, as with $H_W$, the approximant Hamiltonian weakly couples between states that are uncoupled in the quasicrystalline system; see Appendix \ref{App:Approximant_Symmetries} for further details. This opens gaps of $\sim 10^{-5} E_R$ and $\sim10^{-7}E_R$ respectively for $N_{\rm a}\!=\!3$ and $N_{\rm a}\!=\!7$. 

\begin{figure}[t]
    \centering
    \sidesubfloat[]{\includegraphics[width=0.85\columnwidth]{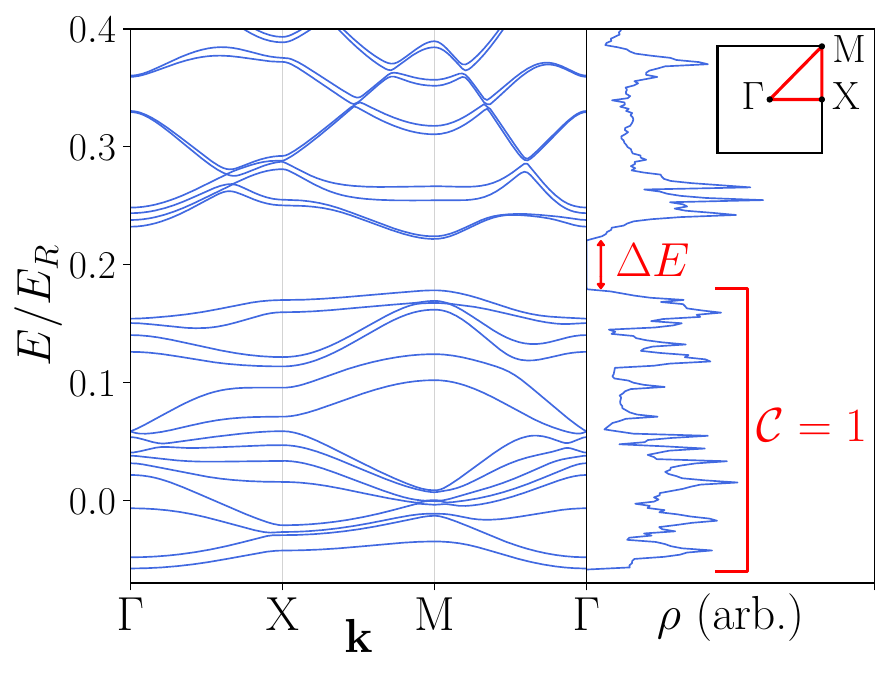}\label{Fig:Approximant_Bandstructure}}\\
    \sidesubfloat[]{\includegraphics[width=0.85\columnwidth]{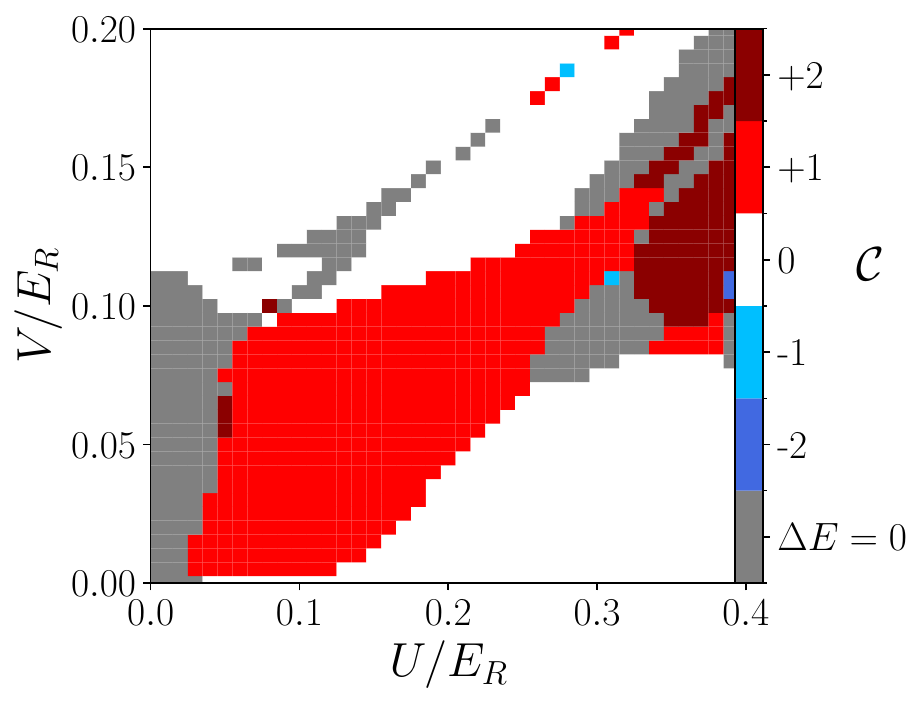}\label{Fig:Phase_Diagram}}
    \caption{(a) Example calculated bandstructure and density of states for the $N_{\rm a}\!=\!3$ approximant system, for $U/E_R\!=\!0.06$ and $V/E_R\!=\!0.02$. The path in the BZ is shown inset, and the energy gap $\Delta E$ is labelled. The set of bands below the gap together have Chern number $\mathcal{C}=1$. (b) Phase diagram showing the Chern number of the bands below the energy gap, for different coupling magnitudes. At grey points, the band gap vanishes and the Chern number is not well-defined.}
    \label{Fig:Approximant_Results}
\end{figure}

Having established that the approximant systems give energy spectra consistent with our quasicrystalline results, we proceed to investigate the topology of our model by analysing approximants throughout the parameter space of $U$ and $V$. Setting $N_{\rm a}\!=\!3$, at each value of $U$ and $V$ we obtain the approximant bandstructure for a mesh of $31\times 31$ points $\mathbf{k}$ in the BZ (our results are converged with respect to increasing the density of the mesh). An example bandstructure is plotted in Fig. \ref{Fig:Approximant_Bandstructure}. We generally find an energy gap $\Delta E$ in the density of states, as predicted from the QBZ analysis. We compute the Berry curvature and Chern number summed over all the bands below the energy gap following the methodology of \cite{Fukui_Chern_2005}. (Due to crossings and touchings of bands, the Chern numbers of the bands individually are not well-defined.) We find the phase diagram depicted in Fig. \ref{Fig:Phase_Diagram}. This displays a large topological region with $\mathcal{C}=1$, confirming our analysis in Sec. \ref{Sec:PlaneWave}, and also demonstrating that the topological properties persist beyond the weak-coupling limit. We note that although our calculations here and in Section \ref{Sec:PlaneWave} are for edge-less geometries, one may naturally expect to find localised boundary modes if the model is implemented in a finite geometry, as with any topological system. In Appendix \ref{App:Trends}, by investigating the variation of $\Delta E$ and the extent of the topological region in parameter space for increasingly accurate approximants (larger $N_{\rm a}$), we show that these topological properties will be preserved in the quasicrystalline limit $N_{\rm a}\to\infty$. 

As Fig. \ref{Fig:Phase_Diagram} shows, for certain coupling strengths the set of bands can attain Chern numbers other than 1 or 0. Unlike the emergence of $\mathcal{C}=1$ in the weak-coupling limit, we do not have a straightforward physical explanation for this behaviour; it may result from higher-order couplings between bands with $\mathcal{C}\neq0$. Additionally, at these coupling strengths, the energy gap is generally small in magnitude ($\Delta E/E_R \lesssim0.01)$, so we cannot be confident that these features will persist from the approximant system to the exact quasicrystal. We stress that this uncertainty relates to the approximation of the quasicrystalline system by the approximant, not a lack of convergence of our calculations for the approximant itself.

\subsection{Filling Density}\label{Sec:Density}

In this section, we calculate the particle density required to fill all states below the band gap in various approximant systems. We expect this to tend to the value obtained from the QBZ area [Eq. (\ref{Eq:Density})] in the limit $N_{\rm a}\!\to\! \infty$, confirming our analysis in Sec. \ref{Sec:PlaneWave} that the isolated band in the quasicrystal is formed by the opening of gaps around the QBZ.

\begin{figure}[t]
    \centering
    \includegraphics[width=0.9\columnwidth]{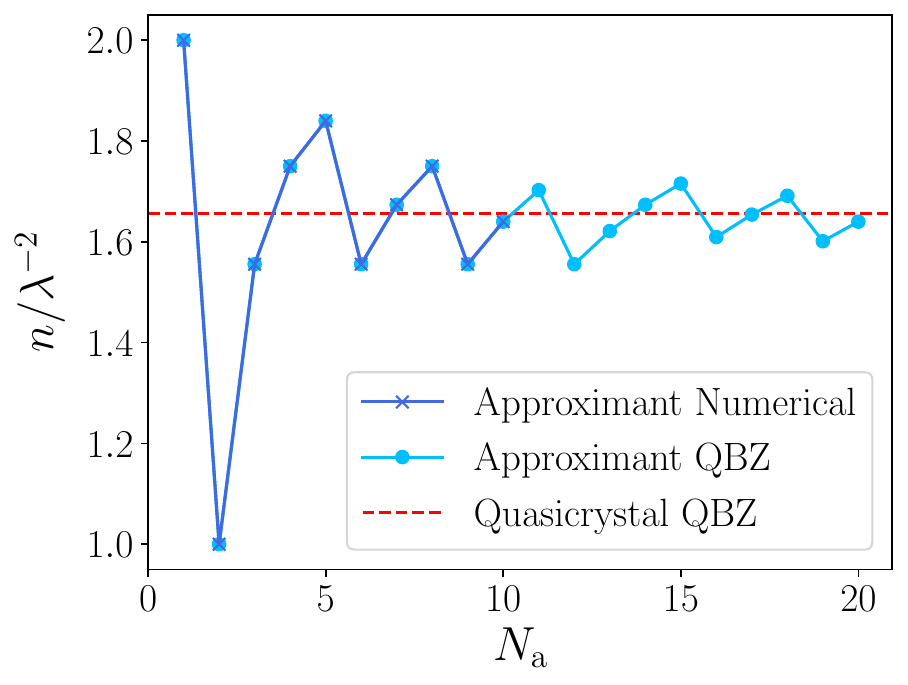}
    \caption{Filling density for increasingly accurate approximants. The quasicrystal value calculated from the QBZ area in Eq. (\ref{Eq:Density}) is shown dashed in red. Dark blue points are calculated from Eq. (\ref{Eq:Approximant_Density}) using the numerically obtained values of $N_\text{B}$, while light blue points are calculated from Eq. (\ref{Eq:Density}) using the approximant QBZ areas, and agree exactly where numerical results are available. For $N_{\rm a}>10$, the full numerical calculations were too expensive to perform; however, our QBZ method can still be applied, and gives results approaching the quasicrystal value as $N_{\rm a}$ increases further.}
    \label{Fig:Filling_Density}
\end{figure}

The approximant densities can be inferred from the numerically calculated bandstructures using:
\begin{equation}\label{Eq:Approximant_Density}
    n=\frac{N_\text{B}A_{\text{BZ}}}{(2\pi)^2},
\end{equation}
where $A_{\text{BZ}}=(|\mathbf{G}|/N_{\rm a})^2$ is the BZ area for the approximant $N_{\rm a}$, and $N_{\rm B}$ is the number of bands below the gap (which is specific to each approximant). The values are plotted in Fig. \ref{Fig:Filling_Density}, and do tend towards the quasicrystalline value as $N_{\rm a}$ increases.

We can further demonstrate the validity of our QBZ analysis by considering the QBZ not of the quasicrystalline system, but of each approximant individually. By the same reasoning as in the quasicrystalline case, energy gaps in the approximant systems will first open along the perpendicular bisectors of the vectors $\widetilde{\mathbf{G}}_l$. This defines the QBZ of the approximant (depicted in Fig. \ref{Fig:Approximant_Diagram}), which is an irregular octagon, unlike the regular octagon of the exact quasicrystal. The area $\widetilde{A}_{\rm QBZ}$ of this region determines the number of states below the gap through Eq. (\ref{Eq:Density}). We expect the densities calculated using this method to agree exactly with those of Eq. (\ref{Eq:Approximant_Density}) for each approximant $N_{\rm a}$ individually, not just in the limit of $N_{\rm a}\!\to\!\infty$. This agreement is evident in Fig. \ref{Fig:Filling_Density}, confirming that the gap we find in the approximant bandstructures is really the same gap that opens on the QBZ boundary. The two (spin-degenerate) free-particle bands over the QBZ are backfolded into $N_{\rm B}$ bands over the BZ, preserving the total number of states below the gap.

\subsection{Narrow Chern Bands}

\begin{figure}[t]
    \centering
    \includegraphics[width=0.9\columnwidth]{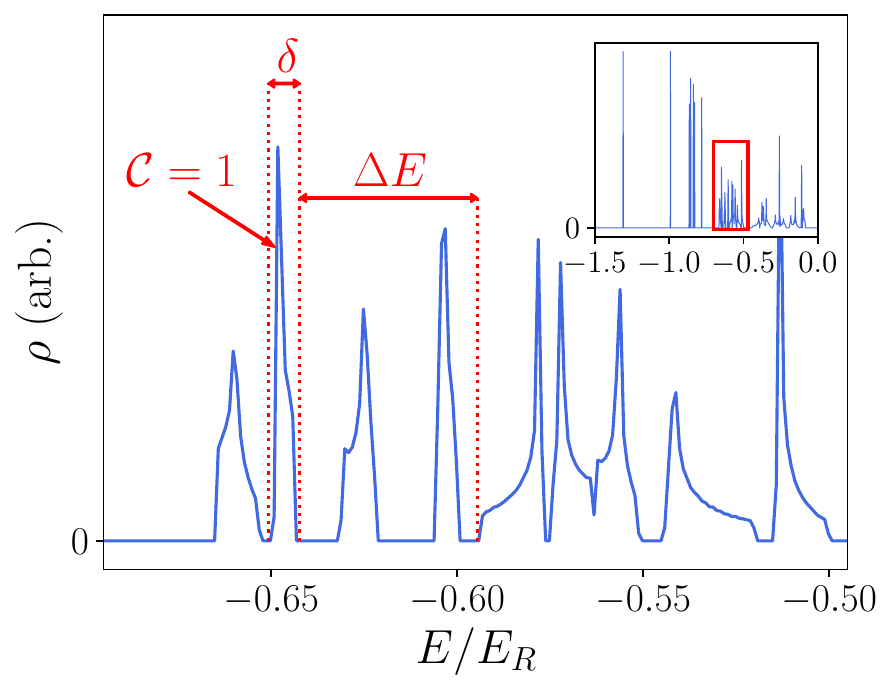}
    \caption{Density of states for an approximant system with $U/E_R=0.3, V/E_R=0.12$. The band with width $\delta$ is topological with $\mathcal{C}=1$. The nearest topologically non-trivial band (with $\mathcal{C}=-1$) is separated by the indicated energy $\Delta E$. The inset shows the density of states over a wider range of $E$; the main plot is the region within the red box. The bands at lower energies are all topologically trivial and irrelevant for our considerations here.}
    \label{Fig:Narrow_Band}
\end{figure}

Our numerical calculations also reveal the potential for realising narrow Chern bands, whose bandwidth $\delta$ is much less than their gap to other topological bands $\Delta E$. (Non-topological bands can be filled and mixed with the topological band without affecting the net properties, so can be ignored.) This could enable strongly-correlated phenomena to be realised in these systems, since interaction effects will dominate over kinetic energy, but not close the band gap. One such example is shown in Fig. \ref{Fig:Narrow_Band}, with a ratio $\delta/\Delta E\approx0.089$. 

The ratio $\delta/\Delta E$ could be further decreased by thoroughly searching for optimal values of $U$ and $V$ in parameter-space, or adding scalar potential terms with the aim of reducing the Hamiltonian to Aharonov-Casher (AC) form following Ref. \cite{Sommer_Ideal_2025}. We have not yet performed any investigations along these lines, but simply note here these possible avenues for future work.

We have attempted to generalise the `dark-state' model discussed in Refs. \cite{Sommer_Ideal_2025, Nascimbene_Emergence_2025} to a quasicrystalline geometry. (See Refs. \cite{Dalibard_Colloquium_2011, Goldman_Lightinduced_2014} for reviews of previous research into dark-state systems with synthetic gauge fields, and Ref.~\cite{Burba_Twodimensional_2025} for other dark state topological lattices.) As this model exactly takes on AC form for a suitable choice of laser phases and in the infinite-potential limit, it seems to be a promising candidate for realising narrow topological bands in a quasicrystalline system. However, without commensurate basis vectors, we find that the magnetic field $B_\text{eff}(\mathbf{r})$ averages to zero, so the AC description is not applicable.

%% file: Summary.tex
\section{Summary and Outlook}\label{Sec:Summary}

We have presented a model for a topological quasicrystalline system that is defined in terms of its momentum-space structure. We have shown how this model could be realised in a cold-atom experiment using two-photon Raman couplings, though similar models may arise in layered 2D electronic materials. Our model is naturally analysed in reciprocal space by considering couplings between plane wave states. Following this methodology, in the weak-coupling limit we have shown analytically on symmetry grounds that the states within the QBZ form an isolated low-energy band with Chern number $\mathcal{C}=1$. This analysis is closely analoguous to that of the  Haldane model -- a semimetallic state converting to a topological band under the breaking of time-reversal symmetry -- but now in a quasicrystalline system.  We have confirmed our analysis with numerical calculations of periodic approximants to the quasicrystal, finding a topological phase with $\mathcal{C}=1$ which persists over a large region in parameter space, and verifying that the number of states below the topological gap corresponds to the QBZ area. We have also found relatively narrow Chern bands for certain choices of parameters, raising the possibility of realising strongly-correlated phases in these systems, if these bands could be flattened further. 

Our work raises questions about the nature of Chern bands in quasicrystalline systems. One such question relates to the localisation of states. As detailed in Sec. \ref{Sec:Intro}, localisation phenomena in non-topological quasicrystals are already the subject of considerable interest; similarly, the behaviour of a (topological) Landau level in the presence of random disorder  has been extensively studied \cite{huckestein_scaling_1995} and continues to attract attention~\cite{loc-zirnbauer,loc-gruzberg,loc-bhaseen,loc-roy}. The case here of localisation in a topological quasicrystalline system may display even richer physics, due to the interplay between topologically-protected extended states, and localised quasicrystalline states. A second question relates to the nature of strongly-correlated phases when particles within a (nearly) flat band interact. Here, due to the quasicrystalline Hamiltonian, we expect a non-uniform and quasicrystalline particle density, which in turn controls the interactions in the system. This could result in fundamentally different behaviour to the familiar fractional quantum Hall or fractional Chern insulator states, which have uniform and periodic particle densities respectively. Investigating these questions in detail is beyond the scope of this paper, but we note them as potential avenues for future work.

%% file: Pentagonal.tex
\section{Pentagonal System}\label{App:Pentagonal}

\begin{figure}[t]
    \centering
    \includegraphics[width=0.9\columnwidth]{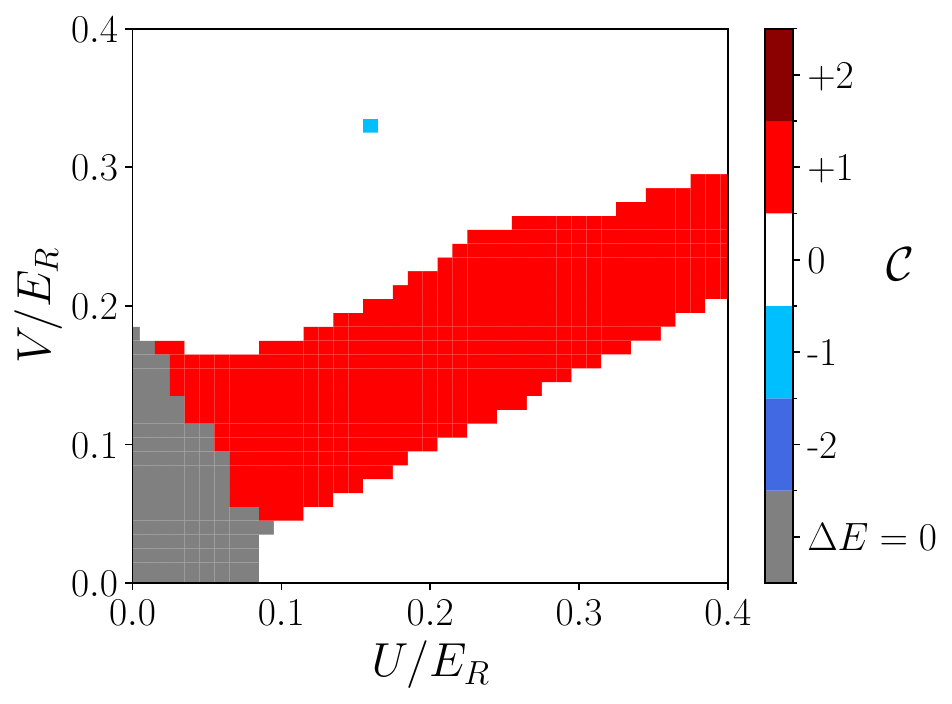}
    \caption{Phase diagram of the 5-fold rotationally symmetric system, calculated for the $N_{\rm a}=4$ approximant.}
    \label{Fig:Phase_Diagram_5fold}
\end{figure}

We have performed the same analysis for an analogous system with 5-fold rotational symmetry. Here, the basis vectors are given by: 
\begin{subequations}
    \begin{equation}
        \mathbf{G}_l=|\mathbf{G}|\,(\cos(2\pi l/5),\,\sin(2\pi l/5))^\text{T}
    \end{equation}
    \begin{equation}
        \mathbf{g}_l=\mathbf{G}_{l+1}-\mathbf{G}_{l}
    \end{equation}
\end{subequations}
with $l\in[0,4]$. We take $V_l=V\!\in\!\mathbb{R}$ as before, and $U$-couplings:
\begin{equation}
    U_l = -U\textrm{e}^{-2\pi \textrm{i} R l/5}
\end{equation}
now with $R=3$. Unlike the previous 8-fold symmetric system, this 5-fold system does not possess symmetry-protected Dirac points at the corners of its (pentagonal) QBZ for $V=0$; instead, it is gapped everywhere along the QBZ boundary by the $U$-couplings. Additionally, the required $U$ to achieve a full spectral gap across the QBZ is larger than for the 8-fold system (due to the larger kinetic energy difference between the $\text{M}$ and $\text{K}$ points here). These factors mean that our plane-wave methodology cannot reliably be applied, since we will violate the requirement $U,\,V\ll E_R$. Instead, we simply analyse the system using periodic approximants. The phase diagram [Fig. \ref{Fig:Phase_Diagram_5fold}] shows that the system becomes topological for $V\sim U$, which is as expected; $V$ must increase until it can close and reopen the non-topological $U$-coupling gap. These calculations also confirm the inapplicability of the plane-wave methodology by explicitly showing that the system is gapless when $U,\,V\ll E_R$. However, interestingly, we have found that the QBZ still accurately predicts the number of states below the gap in the topological phase. 

%% file: Coherent.tex
\section{Coherent Scheme}\label{App:Coherent}

As discussed in the main text, achieving 16 pairs of coherent lasers will pose a significant experimental challenge. Here, we present an alternative scheme requiring only 5 coherent lasers, which may be more realistic in practice. In this alternative scheme, we will choose all lasers to be coherent with each other. As before, we require in-plane lasers with wavevectors $\mathbf{G}_l$, although using reflections means that only four separate lasers are required. Two-photon processes coupling between these lasers will generate all required $H_V$ terms at momenta $\mathbf{g}_l$, provided the electric field vectors $\boldsymbol{\epsilon}^\mathbf{G}_l$ are chosen not directly along $z$, but with some component in the $xy$-plane. Then, we introduce one additional laser with $\sigma_+$ polarisation propagating in the $z$ direction. Two-photon processes between this laser and the in-plane lasers generate the $H_U$ terms.

As indicated in the main text, this scheme comes with two additional complexities: the matrix elements $U_l$ and $V_l$ become dependent on each other, and we get additional couplings between different momentum states, namely at momentum transfers $\mathbf{G}_{l+2}-\mathbf{G}_l$ and $\mathbf{G}_{l+3}-\mathbf{G}_l$. We have analysed this scheme for approximant systems, and still find the same qualitative conclusions: for an extended region of parameter space, the system is a Chern insulator with $\mathcal{C}=1$. 

%% file: Symmetries.tex
\section{Symmetry Analysis}\label{App:Symmetries}

\subsection{Symmetries of the Haldane Model}\label{App:Haldane_Symmetries}

\begin{figure}[t]
    \centering
    \includegraphics[width=0.9\columnwidth]{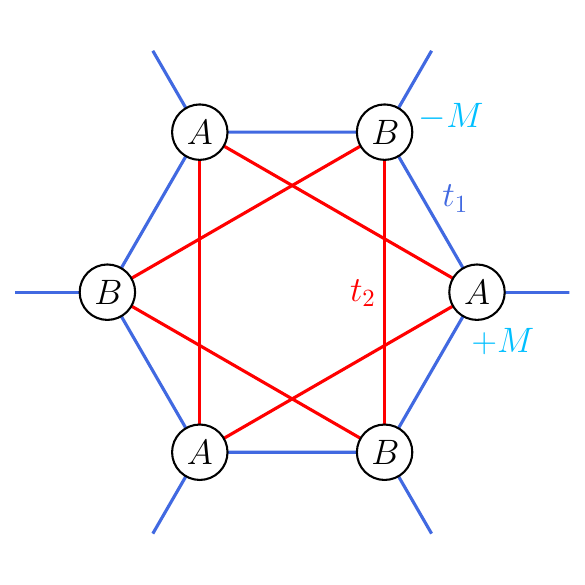}
    \caption{Couplings of the Haldane Model. Sublattice sites $A$ and $B$ lie on a hexagonal lattice, with on-site energies $\pm M$. Particles hop between neighbouring sites with amplitude $t_1$, and next-nearest neighbours with amplitude $t_2$.}
    \label{Fig:Haldane_Model}
\end{figure}

We briefly summarise the symmetries of graphene and the Haldane model \cite{haldane_model_1988} to facilitate comparisons to our model. We first consider graphene. This contains two sites $A$ and $B$ in each unit cell of the hexagonal lattice (see Fig. \ref{Fig:Haldane_Model}), so the Hamiltonian in reciprocal space at crystal momentum $\mathbf{k}$ can be written as a $2\times 2$ matrix in this sublattice basis:
\begin{equation}\label{Eq:GrapheneHamiltonian}
    H_0(\mathbf{k}) =
    \begin{pmatrix}
    0 &
    t_1 \displaystyle\sum_{i=1}^3 {\rm e}^{\rm i \mathbf{k}\cdot \boldsymbol{\delta}_i}
    \\
    t_1 \displaystyle\sum_{i=1}^3 {\rm e}^{-\rm i \mathbf{k}\cdot \boldsymbol{\delta}_i}
    &
    0
    \end{pmatrix} ,
\end{equation}
where, $t_1$ is the hopping amplitude between nearest-neighbours, separated by displacements $\boldsymbol{\delta}_i$. The off-diagonal structure of $H_0(\mathbf{k})$ is enforced by the presence of both inversion symmetry and TRS. Additionally, $C_3$ rotational symmetry between the BZ corners K forces the off-diagonal terms to vanish, resulting in two-fold degenerate Dirac points in the bandstructure.

In the Haldane model, two different symmetry-breaking terms are added to Eq. (\ref{Eq:GrapheneHamiltonian}) which gap out the system. The first is the `Semenoff mass' term:
\begin{equation}
    H_1(\mathbf{k}) = M \sigma_z\, ,
\end{equation}
which breaks inversion symmetry by adding an energy offset $2M\in\mathbb{R}$ between the two sublattices. The second is next-nearest-neighbour hopping with amplitude $t_2 \in \mathbb{C}$, giving terms:
\begin{equation}
    \begin{split}
        H_2(\mathbf{k}) = \: & 2|t_2|\cos(\phi) \sum_{i=1}^3\cos(\mathbf{k}\cdot\mathbf{b}_i) \mathbbm{1}_2 \\
        - & 2|t_2|\sin(\phi) \sum_{i=1}^3\sin(\mathbf{k}\cdot\mathbf{b}_i) \sigma_z\,,
    \end{split}
\end{equation}
where $\phi = \text{arg}(t_2)$, and $\mathbf{b}_i$ are the lattice vectors between sites on the same sublattice. This breaks TRS when $\sin(\phi)\neq0$. Both of these symmetry-breaking terms open a gap at the Dirac point due to their $\sigma_z$ coupling. For $M>3\sqrt{3}\,t_2\sin(\phi)$, the TRS-preserving $H_1$ dominates and the model is a trivial insulator with Chern number $\mathcal{C}=0$; for $M<3\sqrt{3}\,t_2\sin(\phi)$, the TRS-breaking $H_2$ dominates, resulting in $\mathcal{C}=1$.

\subsection{Symmetries of \texorpdfstring{$H_U$}{HU}}

Analysis using group theory of the symmetry operations which commute with the Hamiltonian enable the degeneracies of eigenvalues at different points in the QBZ to be determined. As in Sec. \ref{Sec:QBZ_Bandstructure}, we work in the weak-coupling limit $U,V,W\ll E_R$, and restrict our basis to the 16 states whose kinetic energies are degenerate at the K point. We set this degenerate kinetic energy to be zero without loss of generality, and determine the degeneracies of the eigenspectrum purely by analysing $H_U$, $H_V$ and $H_W$. We emphasise that the following analysis is only valid in this weak-coupling limit where one may neglect coupling to states with different kinetic energies.

\begin{figure}[t]
    \centering
    \includegraphics[width=0.9\columnwidth]{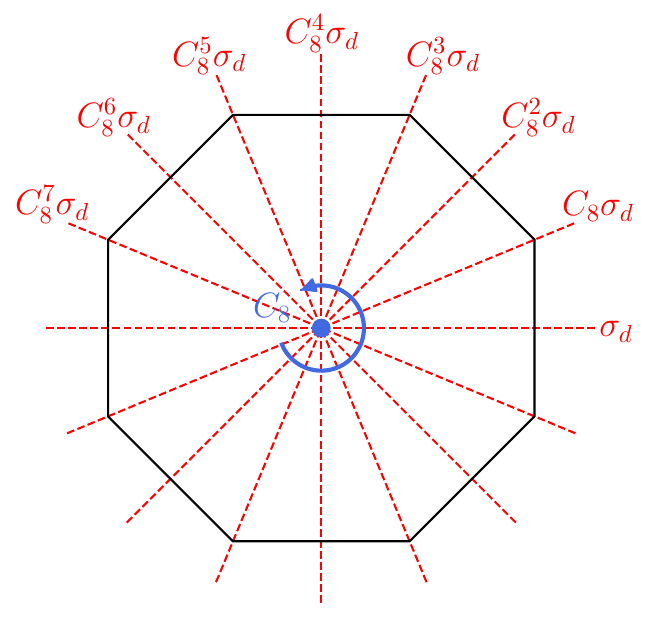}
    \caption{Spatial symmetry operations in the QBZ: an 8-fold rotation axis (blue) and mirror planes (dashed red). The mirror operations can be expressed as combinations of a single mirror plane $\sigma_d$ with powers of the rotation $C_8$.}
    \label{Fig:Symmetries}
\end{figure}

We first analyse $H_U$, which plays the role of $H_0$ in the Haldane model, giving rise to the Dirac points at the QBZ corners K. One symmetry of $H_U$ is the unitary operation:
\begin{equation}
    \widetilde{C}_8 = C_8 \otimes \exp (\textrm{i}\,\Delta\phi\,\sigma_z/2)
\end{equation}
where $C_8$ is an 8-fold spatial rotation in the 2D plane about the origin (see Fig. \ref{Fig:Symmetries}), and $\Delta\phi=\text{arg}(U_{l+1}/U_l)=-2\pi R/8$ is the phase shift between successive $U_l$ matrix elements. A second unitary symmetry is:
\begin{equation}
    \widetilde{\sigma}_d=\sigma_d\otimes \sigma_y
\end{equation}
where $\sigma_d$ is the mirror plane indicated in Fig. \ref{Fig:Symmetries}. Combinations of these operations $\widetilde{C}_8^n$ and $\widetilde{C}_8^n \, \widetilde{\sigma}_d$ are also symmetries. A final symmetry is the anti-unitary time-reversal symmetry:
\begin{equation}
    \mathcal{T}_x=\mathbbm{1}_\mathbf{r}\otimes \sigma_x\mathcal{K}
\end{equation}
where $\mathcal{K}$ denotes complex conjugation. Note that $\mathcal{T}_x^2=+\mathbbm{1}$, which means we do not get any Kramers-type doubling of degeneracies.

We aim to demonstrate that the four-fold degeneracy at the Dirac point is symmetry-protected, rather than accidental. First, we note that the full set of 16 basis states separates into two independent sets of 8 basis states which are not coupled together. The Hamiltonian can therefore be written in block diagonal form:
\begin{equation}
    H = \begin{pmatrix}
        H_1 & 0 \\
        0 & H_2
    \end{pmatrix}
\end{equation}
with $H_1$ and $H_2$ $8\times 8$ matrices in bases $\{\ket{\mathbf{q}_0,\uparrow}, \ket{\mathbf{q}_1,\downarrow}, \ldots\}$ and $\{\ket{\mathbf{q}_0,\downarrow}, \ket{\mathbf{q}_1,\uparrow}, \ldots\}$ respectively. This can be viewed as block-diagonalizing the Hamiltonian into sectors of the symmetry operator:
\begin{equation}
    S = (-1)^N\otimes\sigma_z
\end{equation}
where $N$ is the number $\mathbf{G}_l$ vectors needed to scatter the given momentum state $\mathbf{q}_i$ to $\mathbf{q}_0$. Explicitly, if $\mathbf{q}_i=\mathbf{q}_0 + \sum_l n_l \mathbf{G}_l$, $N=\sum_l |n_l|$. We first consider $H_1$ on its own. $[\widetilde{C}_8^n, S]\neq0$ for odd $n$, since $\widetilde{C}_8$ takes $\mathbf{q}_i\to\mathbf{q}_{i+1}$, which have opposite signs of $(-1)^N$, but leaves spin unchanged. As such, within this sector of $S$, only even powers $\widetilde{C}_8^{2n} = \widetilde{C}_4^n$ remain rotational symmetries. $[\widetilde{\sigma}_d, S]=0$ as $\widetilde{\sigma}_d$ interchanges momentum states with opposite signs of $(-1)^N$ but also flips the spin.

$\widetilde{C}_4$ and $\widetilde{\sigma}_d$ obey the relations:
\begin{subequations}
    \begin{equation}
        \widetilde{\sigma}_d \, \widetilde{C}_4 \, \widetilde{\sigma}_d = (\widetilde{C}_4)^{-1}
    \end{equation}
    \begin{equation}
        \widetilde{C}_4^8 = \mathbbm{1}
    \end{equation}
    \begin{equation}
        \widetilde{\sigma}_d^2=\mathbbm{1}
    \end{equation}
\end{subequations}
and are therefore generators of $D_8$, the group of symmetries of an octagon. (Note that since $R$ is odd, we have $4\Delta\phi=\pi \: (\text{mod} \, 2\pi)$, and therefore $\widetilde{C}_4^4=-\mathbbm{1}$, not simply $\mathbbm{1}$ as might naively be expected.) We can explicitly construct a representation for the symmetry operations as $8\times 8$ matrices which commute with $H_1$. This representation decomposes into irreducible representations $2E_1 \oplus 2E_3$. Since these are all two-dimensional, all eigenvalues of $H_1$ are doubly degenerate.

We now show that $H_1$ and $H_2$ have identical eigenspectra, and therefore all eigenvalues of $H_U$ are four-fold degenerate at the $\text{K}$ point. Firstly, we note that $\{H_U,\mathcal{P}\}=0$, where $\mathcal{P}$ represents spatial inversion, which relies on having $R$ odd so that $U_{l+4}=-U_l$. This means that the spectrum of $H_U$ is symmetric about 0; if $\mathbf{v}$ is an eigenvector with eigenvalue $\lambda$, then $\mathcal{P}\mathbf{v}$ is an eigenvector with eigenvalue $-\lambda$. Secondly, since $H_1$ is Hermitian, all eigenvalues are real, so taking the complex conjugate of the equation $H_1\mathbf{v}=\lambda\mathbf{v}$ gives that $\mathbf{v}^*$ is an eigenvector of $H_1^*$ with eigenvalue $\lambda$. By construction (that is, the choice of bases), $H_2=-H_1^*$, and so the two prior results imply that $\mathcal{P}\mathbf{v}^*$ is an eigenvector of $-H_1^*=H_2$ with eigenvalue $\lambda$. Therefore, all eigenvalues of $H_1$ have a degenerate partner in $H_2$, doubling the degeneracy of the $H_1$ eigenvalues. This proves that the coincident Dirac point we find in the bandstructure of $H_U$ is truly protected by the symmetries of the system, rather than being accidental. We note that similar symmetry-protected Dirac points are also found in quasicrystalline $30^{\circ}$ twisted bilayer graphene \cite{ahn_dirac_2018,yu_dodecagonal_2019,moon_quasicrystalline_2019,crosse_quasicrystalline_2021}.

\subsection{Symmetries of \texorpdfstring{$H_V$}{HV}}

$H_V$ plays the role of the next-nearest-neighbour hopping $H_2$ in the Haldane model, opening a gap at the Dirac point and breaking TRS. When $H_V$ is included, we still do not introduce any coupling between the blocks $H_1$ and $H_2$; that is, $[H_V, S]=0$. $H_V$ also commutes with all of the rotational $\widetilde{C}_8$ symmetries, but breaks the mirror $\widetilde{\sigma}_d$ symmetries. This means that the group of symmetry operations of $H_1$ is reduced from $D_8$ down to $C_8$, the cyclic group of order 8. Since the cyclic group is abelian, it only has one-dimensional irreducible representations, and therefore $H_1$ has non-degenerate eigenvalues in general. As before, $H_1$ and $H_2$ have identical spectra; the proof of this is almost identical to that in the previous section, except that here one must use $\{H_U+H_V,\mathcal{P}\,\widetilde{\sigma}_d\}=0$ in place of simply $\{H_U,\mathcal{P}\}=0$. Therefore, eigenvalues of the full $H$ are doubly-degenerate; however, the 4-fold degeneracy at the Dirac point is lifted, creating a gap. As $H_V$ also breaks TRS (does not commute with $\mathcal{T}_x$), so the bands below the gap can acquire a non-zero Chern number. Note however that TRS was not responsible for protecting the Dirac point, unlike in the Haldane model.

\subsection{Symmetries of \texorpdfstring{$H_W$}{HW}}

$H_W$ is analogous to the Semenoff mass in the Haldane model. $H_W$ commutes with all the symmetries of $H_1$, $\widetilde{C}_4^n$ and $\widetilde{C}_4^n \, \widetilde{\sigma}_d$; it breaks $\widetilde{C}_8$ and its odd powers. Also, importantly, $[H_W, S]\neq0$; $H_W$ couples between basis states in the two sectors, so we need to treat all 16 basis states together. We therefore form a representation of $D_8$ in $16 \times 16$ matrices, which decomposes as $4E_1 \oplus 4E_3$. These two-dimensional irreducible representations imply all eigenvalues are doubly-degenerate, but the four-fold degeneracy at the Dirac point is now lifted. Note that since the $D_{16}$ group, like $D_8$, has only one- and two-dimensional irreducible representations, the coupling between $S$ sectors is sufficient to gap the Dirac point regardless of whether $H_W$ breaks $\widetilde{C}_8$. Since $H_W$ preserves TRS, the bands below the gap must have total $\mathcal{C}=0$. 

\subsection{Approximant Symmetries}\label{App:Approximant_Symmetries}

The approximant Hamiltonians are formed from commensurate $\widetilde{\mathbf{G}}_l$ vectors. We first consider the case $V=0$. As is evident from Fig. \ref{Fig:Approximant_Diagram}, these $\widetilde{\mathbf{G}}_l$ vectors have alternating lengths, and so the $\widetilde{C}_8$ symmetry of the quasicrystalline Hamiltonian is broken down to $\widetilde{C}_4$. Additionally, because the $\widetilde{\mathbf{G}}_l$ are commensurate with each other, the approximant Hamiltonian cannot be block-diagonalised into sectors of $S$; all basis states can be coupled together by a suitable sequence of scatterings. As with $H_W$, this therefore gaps the Dirac points even for $V=0$. However, because these sequences may require many scatterings, the coupling between the basis states in different sectors is extremely weak; the $S$ symmetry is only weakly broken. The resulting gap at the Dirac points is therefore extremely small (for example, $\sim 10^{-5} E_R$ for the $N_{\rm a}=3$ approximant with $U/E_R=0.03$).

%% file: Trends.tex
\section{Trends with \texorpdfstring{$N_{\rm a}$}{Na}}\label{App:Trends}

\begin{figure*}[ht]
    \centering
    \sidesubfloat[]{\includegraphics[width=0.45\textwidth]{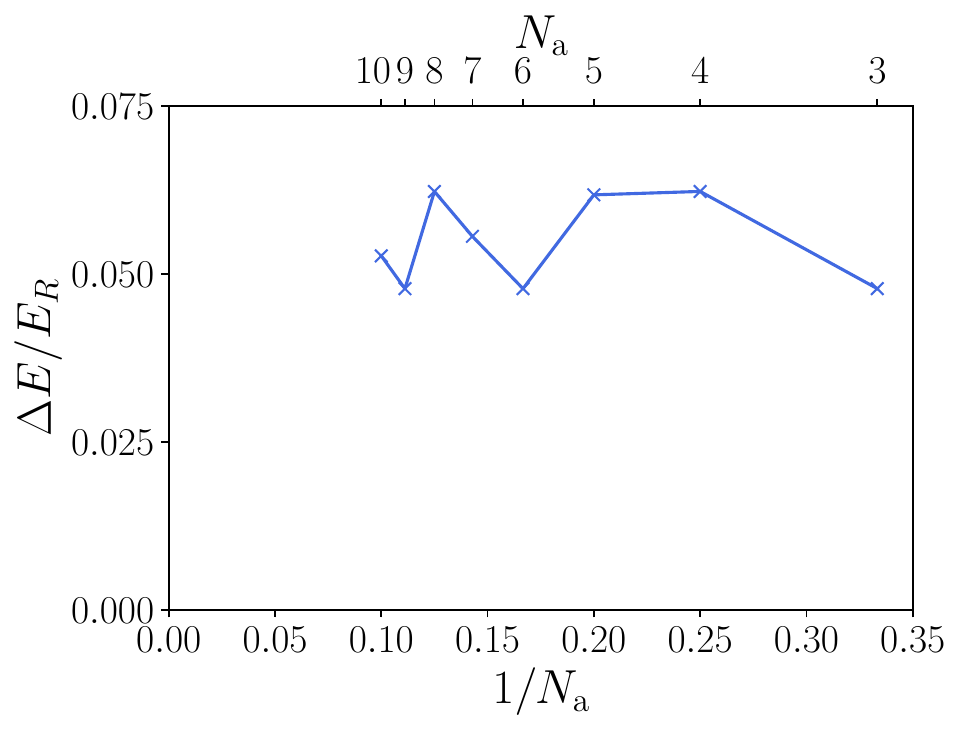}\label{Fig:E_gap_vs_a}}\hfill
    \sidesubfloat[]{\includegraphics[width=0.45\textwidth]{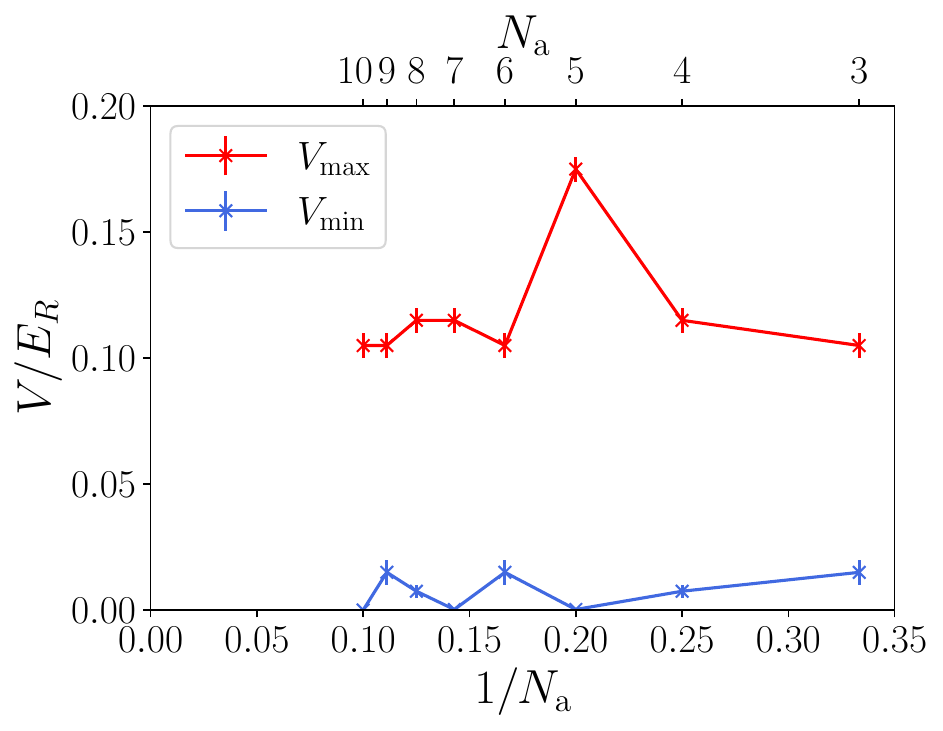}\label{Fig:V_crit_vs_a}}\hfill
    \caption{(a) Energy gap for different approximant systems, for fixed $U/E_R=0.15, V/E_R=0.075$. The quasicrystalline limit is $1/N_{\rm a}\to 0$, that is, the $y$-axis. (b) Upper and lower boundaries of the topological phase for different approximants, for $U/E_R=0.15$. Error bars indicate the confidence interval on each value, generally $\pm0.005$.}
    \label{Fig:Trends}
\end{figure*}

Having established that one approximant system with $N_{\rm a}=3$ is topological over a wide region in parameter-space, we wish to demonstrate that this behaviour will be preserved in the quasicrystalline limit $N_{\rm a}\to\infty$. To do so, we fix the coupling strengths at $U/E_R=0.15, V/E_R=0.075$ (in the centre of the topological region), and calculate the energy gap $\Delta E$ for increasing values of $N_{\rm a}$. Changing $N_{\rm a}$ is simply equivalent to tuning different parameters of the Hamiltonian (the $\widetilde{\mathbf{G}}$-vectors); provided the energy gap does not close under this tuning, the topological index of the bands below the gap cannot change. The results are shown in Fig. \ref{Fig:E_gap_vs_a}. Although there is some variation due to the accuracy of different approximants to the quasicrystal (which does not necessarily improve with every step in $N_{\rm a}$), $\Delta E$ shows no sign of vanishing. 

This is evidence that the quasicrystal will be topological at this single point in parameter space. To go further, we investigate the extent of the topological region in parameter-space for different approximants by fixing $U/E_R=0.15$, and calculating the positions $V_\text{min}$ and $V_\text{max}$ of the lower and upper boundaries of the topological region. The results are plotted in Fig. \ref{Fig:V_crit_vs_a}. Again, there is some variation in the plot, but we can safely conclude that there will be a wide topological region in the quasicrystalline limit.